\documentclass[preprint,secnumarabic,amssymb, nobibnotes, aps, prf, superscriptaddress]{revtex4-2}

\usepackage{amsmath}
\usepackage{graphicx}

\begin{document}

\title{Long-distance migration with minimal energy consumption in a thermal turbulent environment}%

\author{Ao Xu}%
\email[Author to whom correspondence should be addressed: ]{axu@nwpu.edu.cn}
\affiliation{School of Aeronautics, Northwestern Polytechnical University, Xi'an 710072, China}
\affiliation{Institute of Extreme Mechanics, Northwestern Polytechnical University, Xi'an 710072, China}

\author{Hua-Lin Wu}
\affiliation{School of Aeronautics, Northwestern Polytechnical University, Xi'an 710072, China}

\author{Heng-Dong Xi}
\affiliation{School of Aeronautics, Northwestern Polytechnical University, Xi'an 710072, China}
\affiliation{Institute of Extreme Mechanics, Northwestern Polytechnical University, Xi'an 710072, China}

\date{\today}%

\begin{abstract}
We adopt the reinforcement learning algorithm to train the self-propelling agent migrating long-distance in a thermal turbulent environment.
We choose the Rayleigh-B\'enard turbulent convection cell with an aspect ratio ($\Gamma$, which is defined as the ratio between cell length and cell height) of 2 as the training environment.
Our results showed that, compared to a naive agent that moves straight from the origin to the destination, the smart agent can learn to utilize the carrier flow currents to save propelling energy.
We then apply the optimal policy obtained from the $\Gamma=2$ cell and test the smart agent migrating in convection cells with $\Gamma$ up to 32.
In a larger $\Gamma$ cell, the dominant flow modes of horizontally stacked rolls are less stable, and the energy contained in higher-order flow modes increases.
We found that the optimized policy can be successfully extended to convection cells with a larger $\Gamma$.
In addition, the ratio of propelling energy consumed by the smart agent to that of the naive agent decreases with the increase of $\Gamma$, indicating more propelling energy can be saved by the smart agent in a larger $\Gamma$ cell.
We also evaluate the optimized policy when the agents are being released from the randomly chosen origin, which aims to test the robustness of the learning framework, and possible solutions to improve the success rate are suggested.
This work has implications for long-distance migration problems, such as unmanned aerial vehicles patrolling in a turbulent convective environment, where planning energy-efficient trajectories can be beneficial to increase their endurance.
\footnote{
This article may be downloaded for personal use only.
Any other use requires prior permission of the author and APS.
This article appeared in Xu \emph{et al.}, Phys. Rev. Fluids \textbf{8}, 023502 (2023) and may be found at \url{https://doi.org/10.1103/PhysRevFluids.8.023502}.
}
\end{abstract}

\maketitle

\section{Introduction}
Humans have long been fascinated with flight, and we can learn how to fly efficiently from birds.
Some soaring birds can fly long distances during their trips without flapping their wings, and they spend the greatest effort only during the take-off or landing stage.
For example, Weimerskirch \emph{et al.} \cite{weimerskirch2016frigate} showed that frigate birds can stay aloft for up to 48 days during transoceanic flight.
Williams \emph{et al.} \cite{williams2020physical} recorded that an Andean condor flew for over 5 h without flapping, which covers 172 km.
Croxall \emph{et al.} \cite{croxall2005global} revealed that the fastest gray-headed albatrosses can make global circumnavigations in just 46 days.
It was not until 1885, when Lancaster published his pioneer observations and deductions \cite{lancaster1885problem}, that the mystery of flying birds not flapping their wings was gradually solved.
The secret of birds is that they can utilize warm rising atmospheric currents (also known as \emph{thermals}) to reduce the expenditure of energy.
Thermals are part of the convection flows that develop in the convective layer of the atmosphere (i.e., the troposphere).
During sunny days, heat from the sun warms the earth and the earth warms the air above it.
Warm air expands and lighter air rises, and the resulting column of rising air is called \emph{thermals}.

Not only birds but also gliders and unmanned aerial vehicles (UAVs) can utilize the updrafts of thermals to increase endurance and save energy.
For example, MacCready \cite{maccready1958optimum} determined the optimal gliding speed to fly between thermals to maximize speed and energy gain.
Allen \emph{et al.} \cite{allen2005autonomous} estimated a UAV with a nominal endurance of 2 h can achieve a 12-h increase in the summer and a 6-h increase in the winter.
To investigate the use of the convective lift, various thermal models have been developed, such as chimney models and bubble models \cite{bencatel2013atmospheric}.
Chimney thermals are continuous columns of rising air, which extend from the ground surface to the highest level of the troposphere \cite{allen2006updraft}.
Bubble thermals are closed updraft masses that form a rising vortex ring near the ground, and the updraft at the core of the vortex ring is provided by the buoyancy of the air \cite{lawrance2009wind}.
When the air leaves the bubble core, it cools down and loses buoyancy, thus moving downward on the outside of the vortex ring to complete a cycle.
Both the chimney and bubble thermal models describe simplified situations, where there is no turbulent motion or the fluctuations are modeled as Gaussian white noise.
However, in the troposphere, the wind field exhibits strong turbulent fluctuations.
Akos \emph{et al.} \cite{akos2010thermal} found that turbulent fluctuations of the environment bring challenges in identifying effective thermal soaring strategies.
Laurent \emph{et al.} \cite{laurent2021turbulence} further pointed out that turbulence leaves an imprint on all modes of flight, and they revealed the analogy between the flight trajectories of a golden eagle and the trajectories of particles carried by turbulent flows.
They also reinforced the need to fully incorporate turbulence into understanding the movement and behaviors of the flying object.

To model the flow patterns of the wind in strong convective weather, a paradigmatic turbulent convection system, known as Rayleigh-B\'enard (RB) convection, can describe turbulent flows driven by buoyancy forces \cite{ahlers2009heat,chilla2012new,xia2013current}.
The control parameters of the RB system mainly include the Prandtl number (Pr, defined later in the paper), the Rayleigh number (Ra), and the cell aspect ratio ($\Gamma$).
The Pr describes the thermophysical properties of the fluid and Pr = 0.71 for the air.
The Ra describes the ratio of buoyancy forces relative to the viscous forces due to temperature differences, and $10^{18} \le$ Ra $\le 10^{22}$ in the atmosphere \cite{chilla2012new}.
The $\Gamma$ characterizes the geometric information of the convection system, and $\Gamma \approx 100$ for mesoscale convective system \cite{atkinson1996mesoscale}.
In the RB turbulent convection, important coherent structures include small-scale thermal plumes, large-scale circulation rolls, and the very-large-scale superstructure.
The thermal plumes are detached from the hot or cold boundary layers, it then collides and merges, further self-organize into large-scale circulation rolls.
If the convection system extends several times the distance in the horizontal direction than that in the vertical direction, then thermal plumes form a web of connected ridgelike structures of cold downwelling and hot upwelling fluids, also known as the superstructure of thermal turbulence \cite{stevens2018turbulent,pandey2018turbulent}.

Adopting the RB turbulent environment, Reddy \emph{et al.} \cite{reddy2016learning} numerically trained a glider to rise on thermals using reinforcement learning algorithms.
The trained glider can ascend from low altitude to high altitude in a spiral form, which has a similar pattern to soaring birds in nature \cite{akos2008comparing}.
They analyzed the changes in the glider's flight strategy when the turbulent intensity varied.
Thereafter, they equipped a glider with a 2-m wingspan and trained the glider in the field to navigate atmospheric thermals autonomously \cite{reddy2018glider}.
In Reddy \emph{et al.}'s works \cite{reddy2016learning,reddy2018glider}, the main goal is to train the glider to ascend higher;
while for practical application of UAVs, a more frequently encountered scenario is to fly from one position to another.
To minimize energy consumption during the point-to-point migration in a thermal turbulent environment, Xu \emph{et al.} \cite{xu2022migration} optimized the trajectory for a self-propelling agent in the RB turbulent convection, such that the agent can utilize the kinetic energy of the thermal turbulence as much as possible.
Compared with the straight-line propelling trajectory, the optimized trajectory allows the agent to save around two-thirds of its energy consumption.

In the previous work on soaring within the RB turbulent environment \cite{reddy2016learning,xu2022migration}, the simulated RB convection cells have an aspect ratio of $\Gamma \le 2$.
However, migration often occurs in a large-aspect-ratio convection system and covers a long distance.
In this work, our motivation is to train the self-propelling agent to migrate in a large-aspect-ratio RB cell that has multiple circulation rolls.
The rest of this paper is organized as follows.
In Sec. \ref{sec:simulation}, we introduce numerical details for the simulation of the turbulent environment, including the mathematical model and the in-house numerical solver for the RB convection.
In Sec. \ref{sec:dynamics}, we present details of the dynamics of the self-propelling agent and the reinforcement learning algorithm to train the agent to find an energy-efficient trajectory.
In Sec. \ref{sec:results}, we first present general flow patterns in the RB convection, followed by training results for the agent migrating in a $\Gamma =2$ cell, and then test the agent migrating in larger $\Gamma$ cells.
In Sec. \ref{sec:conclusions}, the main findings of this work are summarized.

\section{Simulation of the turbulent environment \label{sec:simulation}}
\subsection{Mathematical model for the RB turbulent thermal convection}
We simulate the turbulent environment in the RB convection cells based on the Boussinesq approximation.
We assume the fluid flow is incompressible, and we treat the temperature as an active scalar that influences the velocity field through the buoyancy.
The viscous heat dissipation and compression work are neglected, and all the transport coefficients are assumed to be constants.
Then, the governing equations for the RB thermal convection can be written as
\begin{equation}
\nabla \cdot \mathbf{u} = 0,
\label{Eq.continuity}
\end{equation}
\begin{equation}
\frac{\partial \mathbf{u}}{\partial t}+\mathbf{u}\cdot\nabla \mathbf{u}
=-\frac{1}{\rho_{0}}\nabla P
+\nu \nabla^{2}\mathbf{u}
+g\beta_{T}(T-T_{0})\hat{\mathbf{y}},
\label{Eq.momentum}
\end{equation}
\begin{equation}
\frac{\partial T}{\partial t}+\mathbf{u}\cdot \nabla T=\alpha_{T}\nabla^{2}T,
\label{Eq.energy}
\end{equation}
where $\mathbf{u}=(u,v)$, $P$ and $T$ are the velocity, pressure, and temperature of the fluid, respectively.
$\rho_{0}$ and $T_{0}$ are reference density and temperature, respectively.
$\hat{\mathbf{y}}$ is the unit parallel to gravity.
With the scaling
\begin{equation}
    \begin{split}
& \mathbf{x}^{*} = \mathbf{x}/H, \ \ t^{*}=t/\sqrt{H/(\beta_{T}g\Delta_{T})}, \ \ \mathbf{u}^{*}=\mathbf{u}/\sqrt{\beta_{T}gH\Delta_{T}}, \\
& P^{*}=P/(\rho_{0}g\beta_{T}\Delta_{T}H), \ \ T^{*}=(T-T_{0})/\Delta_{T}.
    \end{split}
\end{equation}
Then, Eqs. (\ref{Eq.continuity})-(\ref{Eq.energy}) can be rewritten in dimensionless from as
\begin{equation}
\nabla \cdot \mathbf{u}^{*} = 0,
\end{equation}
\begin{equation}
\frac{\partial \mathbf{u}^{*}}{\partial t^{*}}
+\mathbf{u}^{*}\cdot \nabla \mathbf{u}^{*}
=-\nabla P^{*}+\sqrt{\frac{Pr}{Ra}}\nabla^{2}\mathbf{u}^{*}
+T^{*}\tilde{\mathbf{y}},
\end{equation}
\begin{equation}
\frac{\partial T^{*}}{\partial t^{*}}+\mathbf{u}^{*}\cdot \nabla T^{*}
=\sqrt{\frac{1}{Pr Ra}}\nabla^{2}T^{*}.
\end{equation}
Here, $H$ is the cell height and it is chosen as the characteristics length;
$t_{f}=\sqrt{H/(\beta_{T}g\Delta_{T})}$ is the free-fall time and it is chosen as the characteristic time.
$\Delta_{T}$ is the temperature difference between heating and cooling walls.
The two dimensionless parameters are Ra and the Pr, which are defined as
\begin{equation}
\text{Ra}=\frac{g\beta_{T}\Delta_{T}H^{3}}{\nu \alpha_{T}}, \ \ \ \text{Pr}=\frac{\nu}{\alpha_{T}}.
\end{equation}

\subsection{The lattice Boltzmann method for thermal convection}
We adopt the lattice Boltzmann (LB) method to simulate thermal convection.
The advantages of the LB method include easy implementation and parallelization as well as high computing efficiency \cite{xu2017lattice}.
Specifically, we chose a D2Q9 model for the Navier–Stokes equations to simulate fluid flows and a D2Q5 model for the energy equation to simulate heat transfer.
To enhance the numerical stability, the multi-relaxation-time collision operator is adopted in the evolution equations of both density and temperature distribution functions.
The evolution equation of the density distribution function is written as
\begin{equation}
f_{i}(\mathbf{x}+\mathbf{e}_{i}\delta_{t}, t+\delta_{t})-f_{i}(\mathbf{x}, t)
=-\left(\mathbf{M}^{-1}\mathbf{S} \right)_{ij}\left[\mathbf{m}_{j}(\mathbf{x},t)-\mathbf{m}_{j}^{(\text{eq})}(\mathbf{x},t) \right]
+\delta_{t}F_{i}',
\label{Eq.lbm-f}
\end{equation}
where $f_{i}$  is the density distribution function,
$\mathbf{x}$ is the fluid parcel position,
$t$ is the time,  $\delta_{t}$ is the time step,
and $\mathbf{e}_{i}$ is the discrete velocity along the $i$th direction.
The forcing term $F_{i}'$ on the right-hand side of Eq. (\ref{Eq.lbm-f}) is given by  $\mathbf{F}'=\mathbf{M}^{-1}\left( \mathbf{I}-\mathbf{S}/2 \right)\mathbf{M}\tilde{\mathbf{F}}$,
and the term  $\mathbf{M}\tilde{\mathbf{F}}$ is given as \cite{guo2008analysis}
$\mathbf{M}\tilde{\mathbf{F}}=
\left[0, \ 6\mathbf{u}\cdot\mathbf{F}, \ -6\mathbf{u}\cdot\mathbf{F}, \ F_{x}, \ -F_{x}, \ F_{y}, \ -F_{y}, \
2uF_{x}-2vF_{y}, \ uF_{x}+vF_{y} \right]^{T}$,
where $\mathbf{F}=\rho g \beta_{T}(T-T_{0})\hat{\mathbf{y}}$.
The macroscopic density $\rho$ and velocity $\mathbf{u}$ are obtained from $\rho=\sum_{i=0}^{8}f_{i}$,
$\mathbf{u}=\left(\sum_{i=0}^{8}\mathbf{e}_{i}f_{i}+\mathbf{F}/2 \right)/\rho$.

The evolution equation of the temperature distribution function is written as
\begin{equation}
g_{i}(\mathbf{x}+\mathbf{e}_{i}\delta_{t}, t+\delta_{t})-g_{i}(\mathbf{x}, t)
=-\left(\mathbf{N}^{-1}\mathbf{Q} \right)_{ij}\left[\mathbf{n}_{j}(\mathbf{x},t)-\mathbf{n}_{j}^{(\text{eq})}(\mathbf{x},t) \right],
\end{equation}
where $g_{i}$  is the temperature distribution function.
The macroscopic temperature $T$ is obtained from $T=\sum_{i=0}^{4} g_{i}$.
More numerical details of the LB method and validation of the in-house solver can be found in our previous work  \cite{xu2017accelerated,xu2019lattice,xu2023multi}.

\subsection{Simulation settings for the turbulent thermal convection}
We consider a two-dimensional RB cell with length $L$ and height $H$.
The top and bottom walls of the cell are kept at constant cold and hot temperatures, respectively;
while the other two vertical walls are adiabatic; all four walls impose no-slip velocity boundary conditions.
We set the cell aspect ratio ($\Gamma=L/H$) as $2 \le \Gamma \le 32$, and we fix the Prandtl number as Pr = 0.71 (corresponds to the working fluids of air) and the Rayleigh number as Ra $= 10^{8}$.
Although the Ra is far less than that in the atmosphere because of limited computing resources to simulate ultra-high Ra convection, we note the RB convection at $\text{Ra} = 10^{8}$ already exhibits strong turbulent fluctuations and the flows fall in the "hard turbulence" regime \cite{castaing1989scaling}.
We also checked the turbulent database and confirmed that statistically stationary states have been reached and the initial transient effects of the simulations are washed out.

\section{Dynamics and control of the self-propelling agent \label{sec:dynamics}}
\subsection{Kinematic model of the self-propelling agent}
The dynamics of the self-propelling agent can be described as
\begin{equation}
\mathbf{u}_{\text{agent}}(t)=\mathbf{u}_{\text{fluid}}(t)+\mathbf{u}_{\text{propel}}(t)
\end{equation}
\begin{equation}
\mathbf{x}_{\text{agent}}(t+dt)=\mathbf{x}_{\text{agent}}(t)+\mathbf{u}_{\text{agent}}(t)\cdot dt
\end{equation}
Here $dt$ is the time step, $\mathbf{u}_{\text{agent}}$ and $\mathbf{x}_{\text{agent}}$ denote the velocity and position of the agent, respectively.
$\mathbf{u}_{\text{fluid}}$ denotes the velocity of the carrier fluids, and $\mathbf{u}_{\text{propel}}$  denotes the velocity generated by the agent.
We assume that, without control, the velocity of the agent equals that of the carrier flows;
while, with control, the velocity of the agent is the superposition of the carrier flow and agent's propulsion.
Similar dynamics of the agent have been previously adopted by Krishna \emph{et al.} \cite{krishna2022finite}.
To mimic the limited propelling ability of the agent in real-world scenarios,
we restricted the maximum propelling velocity of the agent $\|\mathbf{u}_{\text{agent}}\|$ to be less than one-third of the largest carrier flow velocity.
On the other hand, a more complex kinematic model for the self-propelling agent, such as the one that includes inertial and rotational dynamics \cite{colabrese2017flow,colabrese2018smart,schneider2019optimal,alageshan2020machine,borra2022reinforcement}, fluttering and tumbling \cite{novati2019controlled}, and multimodal locomotion \cite{zou2022gait} of the propelling agent, can be considered in the future work.

\subsection{Optimal control via the reinforcement learning}
We adopt the RL algorithm to optimize the control of the agent to migrate in an energy-efficient trajectory.
The advantages of the RL algorithm include agnostic for control and optimization tasks, easy to be re-used to speed-up optimization in a similar system configuration, and robust to the disturbances in the chaotic system  \cite{biferale2019zermelo,garnier2021review}.
In the RL algorithm, the agent observes the state of the environment and decides to take an action interacting with the environment.
If the agent then receives a reward (or a penalty), then it is more likely to repeat (or forego) that action in the future.
Overall, the agent learns by trial and error, with the long-term goal to maximize the cumulative expected return and eventually improve its performance.
Applying the RL algorithm, we can obtain the optimal policy, which advises the favorable action to take for the agent   \cite{sutton2018reinforcement,mehta2019high,brunton2020machine,cichos2020machine}.
The model-free RL algorithm can generally be classified into policy-based methods and value-based methods.
In the policy-based method, such as the policy gradient method, the parameter of the policy network  $\theta$ is optimized to maximize the performance objective $J(\pi_{\theta})$.
Here $\pi_{\theta}$ denotes the parameterized stochastic policy.
The policy-based methods are inefficient in sampling, thus leading to slow learning, and are not suitable for complex flow problems \cite{tsang2020self}.
In the value-based methods, such as the $Q$-learning method, the agent takes action $a$ that tried to maximize the optimal action-value function, i.e.,  $a(s)=\arg \max_{a}Q_{\theta}(s,a)$.
Here $s$ denotes the state of the environment and $Q_{\theta}(s,a)$ approximates the optimal action-value function $Q^{*}(s,a)$.
Using the $Q$-learning method, Colabrese \emph{et al.} \cite{colabrese2017flow} showed that gravitactic swimmers can reach high altitudes in steady Taylor-Green vortex flow,
Mui\~nos-Landin \emph{et al.} \cite{muinos2021reinforcement} demonstrated the artificial self-thermophoretic microswimmers can navigate under the influence of Brownian motion,
Monderkamp \emph{et al.} \cite{monderkamp2022active} trained active Brownian particles through complex motility landscapes,
and Gazzola \emph{et al.} \cite{gazzola2016learning} and Verma \emph{et al.} \cite{verma2018efficient} found optimal swimming strategies that minimize drag and energy consumption in the school of fish.
The above five examples adopt the off-policy learning techniques, which means that each update stochastic samples the data collected at any point during training, namely $Q(s_{t},a_{t})=Q(s_{t},a_{t})+\alpha\left[r_{t+1}+\gamma \max_{a}Q(s_{t+1},a)-Q({s_{t},a_{t}}) \right]$.
Earlier, Reddy \emph{et al.} \cite{reddy2016learning} adopted the state-action-reward-state-action (SARSA) method, which can be regarded as a variation of the $Q$-learning method.
The main difference is that the SARSA method adopts the on-policy learning technique, which uses the action performed by the current policy to learn the $Q$-value, namely $Q(s_{t},a_{t})=Q(s_{t},a_{t})+\alpha\left[r_{t+1}+\gamma Q(s_{t+1},a_{t+1})-Q({s_{t},a_{t}}) \right]$.
However, the value-based method can only work in discrete state and action spaces;
while in most real-world scenarios, such as training a vehicle to navigate, the continuous state and action spaces are preferred to develop more versatile motion for complex navigation tasks.

In this work, we adopt the soft actor-critic (SAC) algorithm, which is an interpolation between policy-based methods and value-based methods.
In the SAC algorithm, the agent decides the next action via the actor network, whilst that action is further evaluated by the critic network to guide the training process.
In addition, the actor aims to maximize the expected reward (i.e., succeed at the task) while also maximizing entropy (i.e., acting as randomly as possible).
In entropy regularized reinforcement learning, the optimization problem can be described as
\begin{equation}
\pi^{*}(\theta)=\arg \max_{\pi} E_{\tau \sim \pi}\left[ \sum_{t=0}^{\infty}\left\{ R(s_{t},a_{t},s_{t+1})+\alpha H[\pi(\cdot | s_{t})] \right\} \right]
\end{equation}
In the above equation,  $\pi^{*}$ is the optimal policy.
The reward function $r$ depends on the current state of the environment $s_{t}$, the current action $a_{t}$, and the next state of the environment $s_{t+1}$.
$\alpha$ is the trade-off coefficient.
The entropy $H$ of $\tau$ is computed from its distribution  $\pi$ as  $H[\pi(\cdot|s_{t})]=E_{\tau \sim \pi}[-\log \pi(\tau)]$.
More details on the SAC algorithm can be found by Haarnoja \emph{et al.} \cite{haarnoja2018soft}.

Key ingredients in the RL framework include the environmental cues that the agent can observe (i.e., the current state $s_{t}$ of the environment), the actions the agent takes (i.e., $a_{t}$), and the response of the agent to its behavior (i.e., the reward $r_{t}$).
In this work, to migrate in a large-aspect-ratio convection cell, the observation variables for the agent include the carrier flow velocity $\mathbf{u}_{\text{fluid}}$, the agent's spatial coordinate in the vertical direction $y_{\text{agent}}$, and the fluid temperature $T$.
In Sec. \ref{sec:traning}, we provide a detailed discussion on selecting observation variables.
The action variable is the propelling velocity $\mathbf{u}_{\text{propel}}$ generated by the agent.
Following the previous work of Xu \emph{et al.} \cite{xu2022migration}, we assume the rewards received by the agent are simultaneously affected by the current state, energy consumption, and time consumption of the agent
\begin{equation}
r(t)=r_{s}(t)+r_{e}(t)+r_{h}(t).
\label{Eq.reward}
\end{equation}
Here the $r_{s}$  denotes the reward affected by the current state of the agent.
If the agent migrates out of the flow domain through the top or the bottom boundaries, then it will receive a penalty of $-\phi$;
if the agent moves rightward and gets closer to the right-side boundary, then it will receive a basic reward $e_{\text{basic}}$ with an empirical prefactor  $\varepsilon$.
Thus, $r_{s}$ is written as
\begin{equation}
r_{s}(t)=\left\{
    \begin{aligned}
& -\phi, \ &  \text{agent is out of the flow domain}  \\
& \varepsilon e_{\text{basic}}, \ & x_{\text{agent}}^{t}>x_{\text{agent}}^{t-1} \\
& 0, \ &  \text{otherwise}  \\
    \end{aligned}
\right.
\label{Eq.reward-rs}
\end{equation}
Here we adopt $\phi=10$ and $\varepsilon=10$.
A detailed discussion on the sensitivity of the hyperparameters in the reward function can be found in Appendix \ref{sec:sensitivity}.
In Eq. (\ref{Eq.reward}), the $r_{e}$ denotes the reward affected by the energy consumption of the agent.
If the propelling velocity of the agent $\mathbf{u}_{\text{propel}}$ is in alignment with that of the background flow $\mathbf{u}_{\text{fluid}}$, namely
the angle between these two vectors (denoted by $\theta$) is less than 90$^{\circ}$,
then the agent will receive a reward of  $\varepsilon[e_{\text{basic}}+(e_{\max}-e)]$, where  $e=0.5||\mathbf{u}_{\text{propel}}||^{2}$ and $e_{\text{basic}}=e_{\max}=0.5(||\mathbf{u}_{\text{propel}}||)^{2}_{\max}$;
if the angle between $\mathbf{u}_{\text{propel}}$ and $\mathbf{u}_{\text{fluid}}$ is greater than 90$^{\circ}$, then the agent will receive a penalty of  $-2\varepsilon(e_{\text{basic}}+e)$.
The above designs also imply that, when the agent migrates in alignment with the carrier flow direction, it will receive a lower reward if the propelling velocity is higher;
when the agent migrates against the carrier flow direction, it will receive a higher penalty if the propelling velocity is higher.
Thus, $r_{e}$  is written as
\begin{equation}
r_{e}(t)=\left\{
    \begin{aligned}
& \varepsilon[e_{\text{basic}}+(e_{\max}-e)], \ &  0^{\circ}\le \theta \le 90^{\circ}  \\
& -2\varepsilon(e_{\text{basic}}+e), \ &  90^{\circ}\le \theta \le 180^{\circ}  \\
    \end{aligned}
\right.
\label{Eq.reward-re}
\end{equation}
In Eq. (\ref{Eq.reward}), the $r_{h}$ denotes the reward affected by the time consumption of the agent.
If the agent migrates out of the domain via the right-side boundary, then we assume the agent completes the task and it will receive a reward that is inversely proportional to the time taken.
This design implies that the sooner the agent reaches the destination, the higher the reward it receives.
Thus, $r_{h}$ is written as
\begin{equation}
r_{h}(t)=\left\{
    \begin{aligned}
& (t_{\max}-t)/\varepsilon, \ &  x_{\text{agent}}^{t}>L  \\
& 0, \ &  \text{otherwise}  \\
    \end{aligned}
\right.
\label{Eq.reward-rh}
\end{equation}

\section{Results and discussion \label{sec:results}}

\subsection{General flow patterns in the RB convection with a large aspect ratio \label{sec:flow}}

\begin{figure}
  \centering
  \includegraphics[width=16cm]{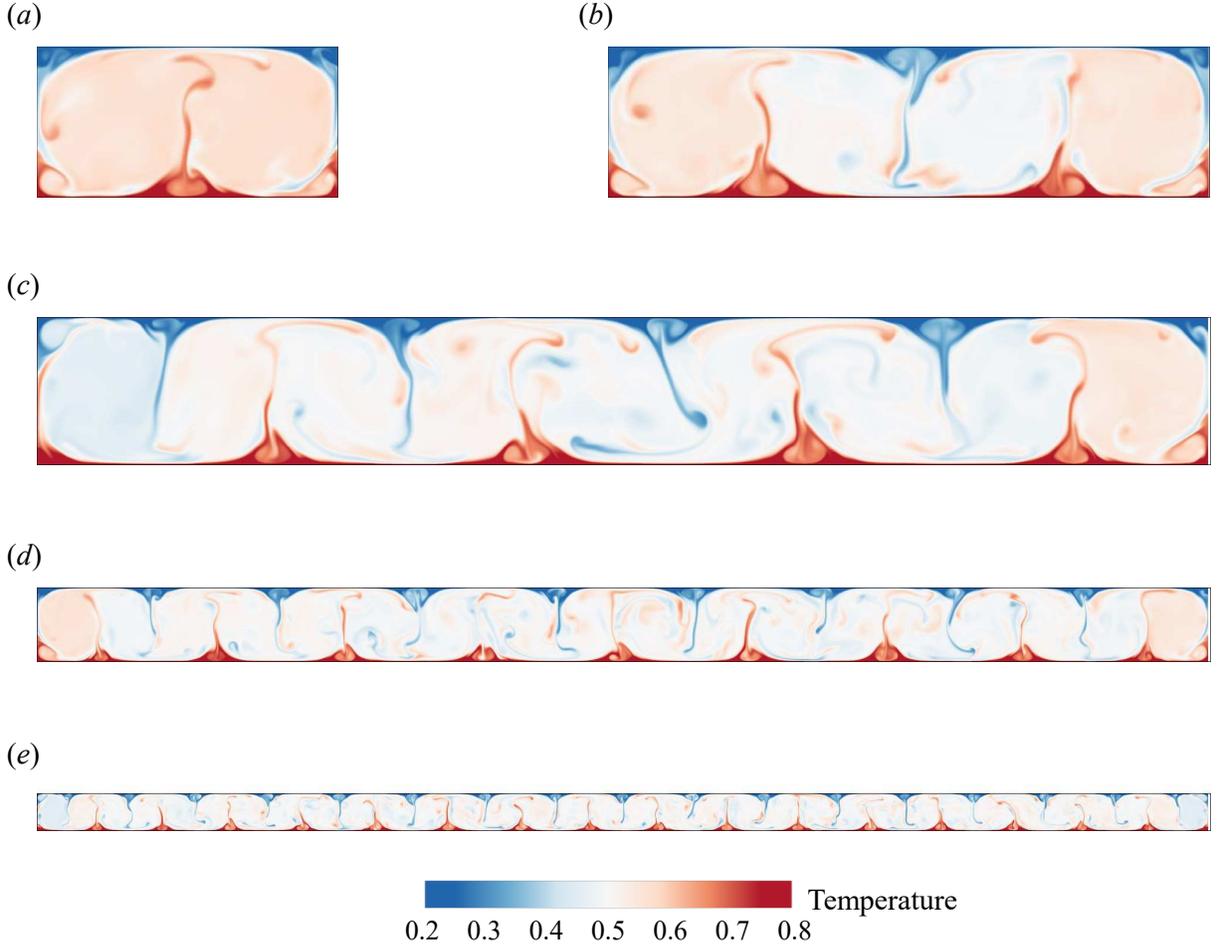}
  \caption{Typical instantaneous temperature field at $\text{Ra} = 10^{8}$, $\text{Pr} = 0.71$,
  (\textit{a}) $\Gamma=2$, (\textit{b}) $\Gamma=4$, (\textit{c}) $\Gamma=8$, (\textit{d}) $\Gamma=16$, and (\textit{e}) $\Gamma=32$. \label{fig:instant-temperature}}
\end{figure}

Typical snapshots of the temperature field at $\text{Ra} = 10^{8}$, $\text{Pr} = 0.71$, and $2 \le \Gamma \le 32$ are shown in Fig. \ref{fig:instant-temperature}.
We can distinguish small-scale thermal plumes and large-scale circulation rolls.
Thin thermal boundary layers appear near the bottom heating wall and top cooling wall.
Plumes that are released from the boundary layers penetrate upwards (or downwards) towards the opposite wall of lower (or higher temperature), and intense mixing occurs in the central region.
Thermals are almost periodically released from relatively fixed locations, and neighboring thermals that move in opposite directions entrain the surrounding fluid and self-organize into circulation rolls.
Previously, Wang \emph{et al.} \cite{wang2020multiple} found that depending on the initial conditions, the flow system at a given aspect ratio evolves to different final turbulent states with different roll numbers.
In our work, the convection roll number $n$ is 2, 4, 9, 18 and 34, in the $\Gamma$  = 2, 4, 8, 16, and 32 convection cells, respectively; their corresponding mean aspect ratios are  $\Gamma_{r}=\Gamma/n=$ 1, 1, 0.889, 0.889 and 0.941, which is consistent with that predicted by the elliptical instability theory \cite{wang2020multiple}.

\begin{figure}
  \centering
  \includegraphics[width=14cm]{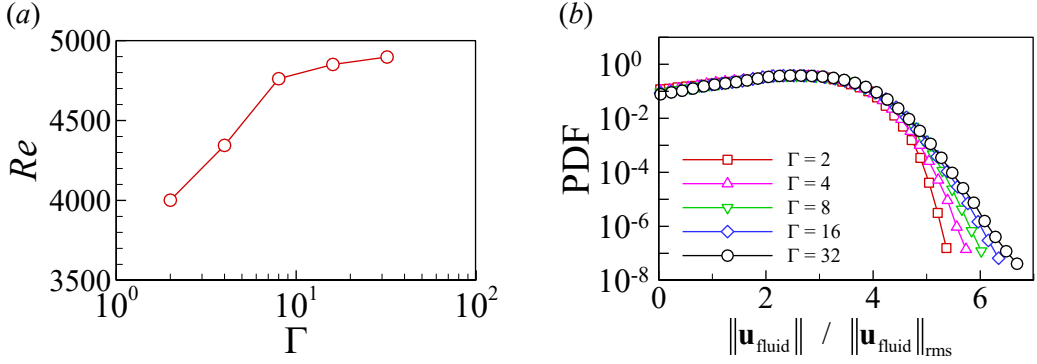}
  \caption{(\textit{a}) Reynolds number
  and (\textit{b}) probability density functions (PDFs) of normalized velocity magnitude
  $\|\mathbf{u}_{\text{fluid}} \|/\|\mathbf{u}_{\text{fluid}} \|_{\text{rms}}$,
  for various $\Gamma$ at $\text{Ra} = 10^{8}$ and $\text{Pr} = 0.71$. \label{fig:Reynolds-PDFu}}
\end{figure}

We then examine the global response parameter of Reynolds number (Re) on the control parameter $\Gamma$.
Here the global flow strength is calculated as  $\text{Re}=\sqrt{\langle (u^{2}+v^{2}) \rangle_{V,t}}H/\nu$ and $\langle \cdots \rangle_{V,t}$  denotes the spatial and temporal average.
The measured Re as a function of $\Gamma$ is shown in Fig. \ref{fig:Reynolds-PDFu}(a), and we can observe enhanced global flow strength with the increase of cell aspect ratio.
In addition, with the increase of $\Gamma$, the Re gradually reaches an asymptotic value, similarly to that in the three-dimensional (3D) convection cell \cite{stevens2018turbulent}.
Previously, Xu \emph{et al.} \cite{xu2022migration} found the optimized energy-efficient strategy obtained from the reinforcement learning algorithm is not sensitive to small perturbation of the global flow strength,
but it changes when the global flow strength increases (or decreases) more than one magnitude of order.
Because the self-propelling agent was trained in the $\Gamma=2$ cell, we further checked that even in the $\Gamma=32$ cell, the flow strength increases around 22\% compared to that in the $\Gamma=2$ cell,
indicating the flow strength in a larger aspect ratio cell does not increase significantly.
To quantify the fluctuations of velocity magnitude, we plot the probability density functions (PDFs) of normalized velocity magnitude $\|\mathbf{u}_{\text{fluid}}\|/\|\mathbf{u}_{\text{fluid}}\|_{\text{rms}}$  for various $\Gamma$, as shown in Fig. \ref{fig:Reynolds-PDFu}(b).
We can see the PDF heads collapse for different $\Gamma$, while the PDF tails become slightly extended with the increase of $\Gamma$, which implies an increased degree of fluctuations for the velocity magnitude  $\|\mathbf{u}_{\text{fluid}}\|$.

\begin{figure}
  \centering
  \includegraphics[width=16cm]{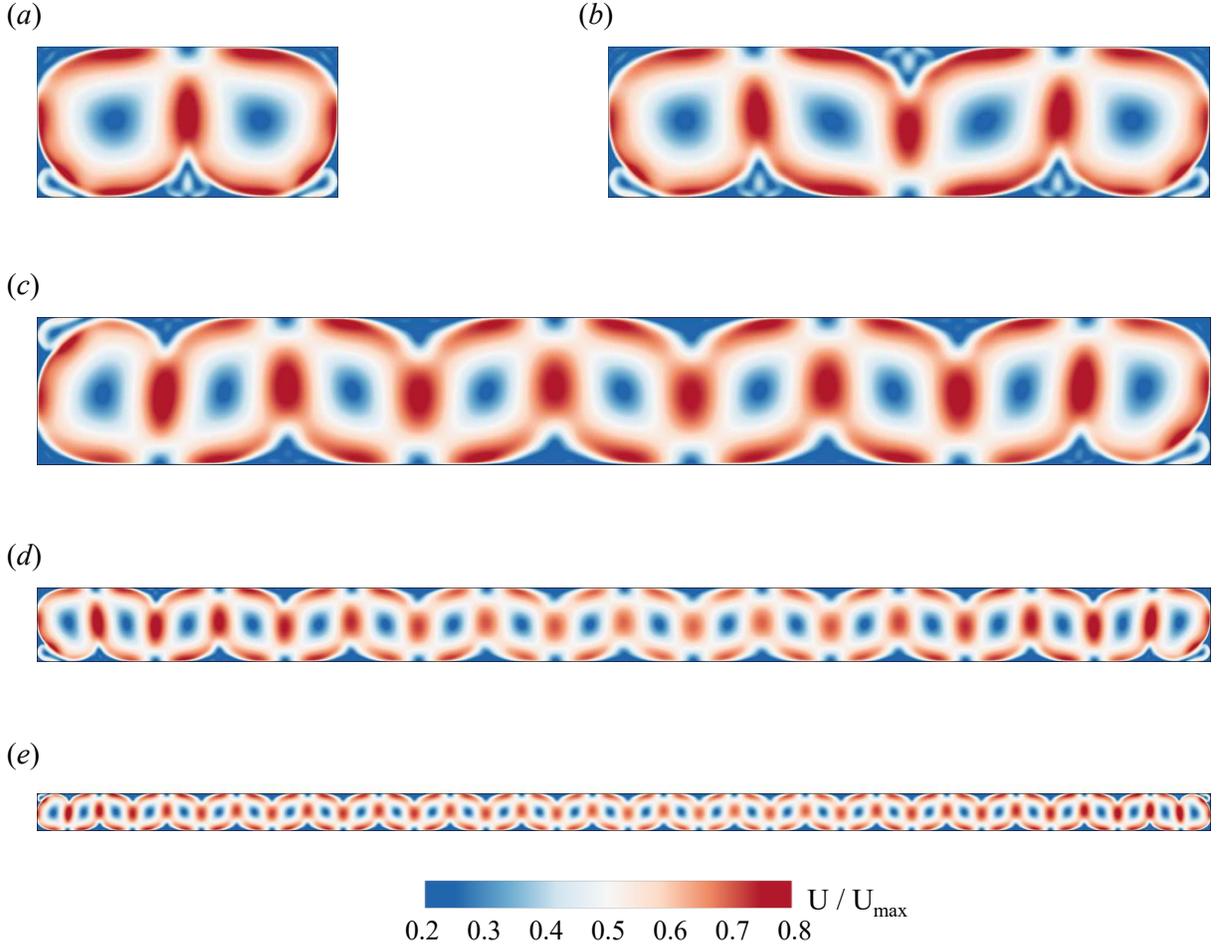}
  \caption{Contour of the first proper orthogonal decomposition (POD) mode $\phi_{1}(\mathbf{r})$
  at $\text{Ra} = 10^{8}$, $\text{Pr} = 0.71$,
  (\textit{a}) $\Gamma=2$, ($\textit{b}$) $\Gamma=4$, ($\textit{c}$) $\Gamma=8$, ($\textit{d}$) $\Gamma=16$, and ($\textit{e}$) $\Gamma=32$.
  Here $U=\sqrt{u^{2}+v^{2}}$  is the velocity magnitude, and $U_{\max}$ denotes the maximum value of $U$ for normalization. \label{fig:POD-mode}}
\end{figure}

To extract the coherent flow structure from the turbulent database, we adopt the proper orthogonal decomposition (POD) analysis.
In the POD, the spatiotemporal vector field $\mathbf{X}(\mathbf{r},t)$  is decomposed as a superposition of orthogonal eigenfunctions $\phi_{i}(\mathbf{r})$  and their amplitudes $a_{i}(t)$  as
\begin{equation}
\mathbf{X}(\mathbf{r},t)=\sum_{i=1}^{\infty}a_{i}(t)\phi_{i}(\mathbf{r})
\end{equation}
Here the vector field  $\mathbf{X}(\mathbf{r},t)$ is chosen as the flow velocity field  $\mathbf{X}=(u,v)$.
Practically, we can use the singular value decomposition (SVD) on the dataset $\mathbf{X}$ to obtain the flow mode $\phi_{i}(\mathbf{r})$  and the corresponding mode amplitude $a_{i}(t)$ \cite{xu2020correlation}.
Because the database for turbulent convection in a large-aspect-ratio cell is huge, which consists of fine spatial resolution and long temporal evolution data, the memory consumption to perform standard SVD is intense.
Here, we adopt the randomized SVD to reduce the computational load \cite{halko2011finding}.
The shape of the first POD mode $\phi_{1}(\mathbf{r})$  at various $\Gamma$ is shown in Fig. \ref{fig:POD-mode}.
The most energetic POD mode consists of horizontally stacked circulation primary rolls rotating in either the clockwise direction or the anticlockwise direction,
and these primary rolls exhibit a periodical pattern.
Large values of velocity magnitude appear near the vortex edge, indicating a strong energy barrier for the agent to move across the vortex edge.
It is noteworthy that at much higher Ra (i.e., $\text{Ra}>10^{10}$), when the large-scale-circulation is weaker and the flow consists of multiple mobile and orbital small vortices \cite{zhang2017statistics,zhu2018transition,wang2020vibration},
whether the current learning framework still works deserves further study.

\begin{figure}
  \centering
  \includegraphics[width=14cm]{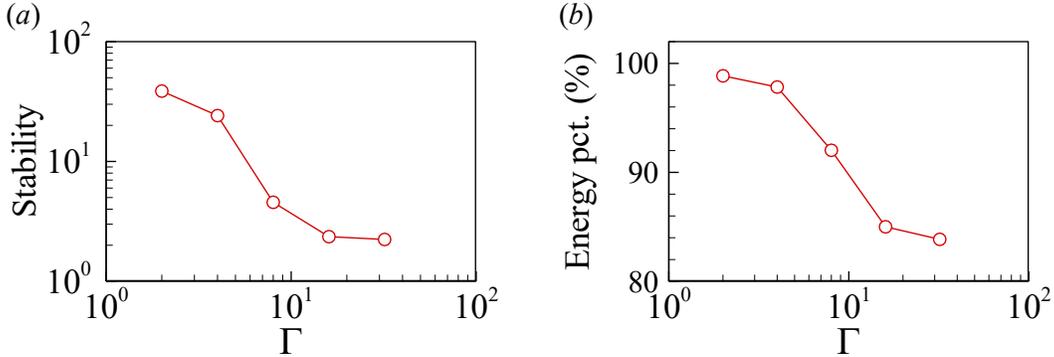}
  \caption{(\textit{a}) The stability of the first POD mode and (\textit{b}) the energy contained in the first POD mode at various $\Gamma$. \label{fig:POD-energy}}
\end{figure}

We further calculate the stability of the first POD mode as $S_{1}=\sqrt{\langle a_{1}(t) \rangle_{t}}/\sigma_{a_{1}}$, such that a larger value of $S_{1}$ indicates a more stable pattern of circulation rolls.
Similar estimations of the roll stability have been previously used in the Fourier mode decomposition of the turbulent thermal convection  \cite{chen2019emergence,xu2020correlation}.
From Fig. \ref{fig:POD-energy}(a), we can see the stability of the first POD mode decreases with the increase of $\Gamma$.
We also analyze the energy contained in the first POD mode and calculate the energy percentage as  $\lambda_{1}/\sum_{i=1}^{\infty}\lambda_{i}$.
Here  we have $\lambda_{i}\delta_{ij}=\langle a_{i}(t)a_{j}(t) \rangle_{t}$ and $\lambda_{i}$ denotes the energy of the $i$th POD mode,
$\delta_{ij}$ is the Kronecker symbol, and  $\langle \cdots \rangle_{t}$ denotes the temporal average.
From Fig. \ref{fig:POD-energy}(b), we can see the energy percentage in the first POD mode also decreases with the increase of $\Gamma$.
Although the first mode is still the dominant flow structure (e.g., in the $\Gamma=32$ cell,
the energy contained in the first POD mode accounts for more than 83.8\% of the total energy), higher-order flow modes become stronger with the increase of $\Gamma$.
Thus, in a large-aspect-ratio cell, the horizontally stacked circulation rolls that form a periodical pattern are less stable,
and those higher-order modes lead to a more irregular flow pattern.
Because the agent was trained in the $\Gamma=2$ cell, the above-mentioned complex flow features bring challenges for the agent to identify an energy-efficient trajectory in a larger $\Gamma$ cell.

\subsection{Training the agent to migrate in the RB cell with an aspect ratio of 2 \label{sec:traning}}
\begin{figure}
  \centering
  \includegraphics[width=14cm]{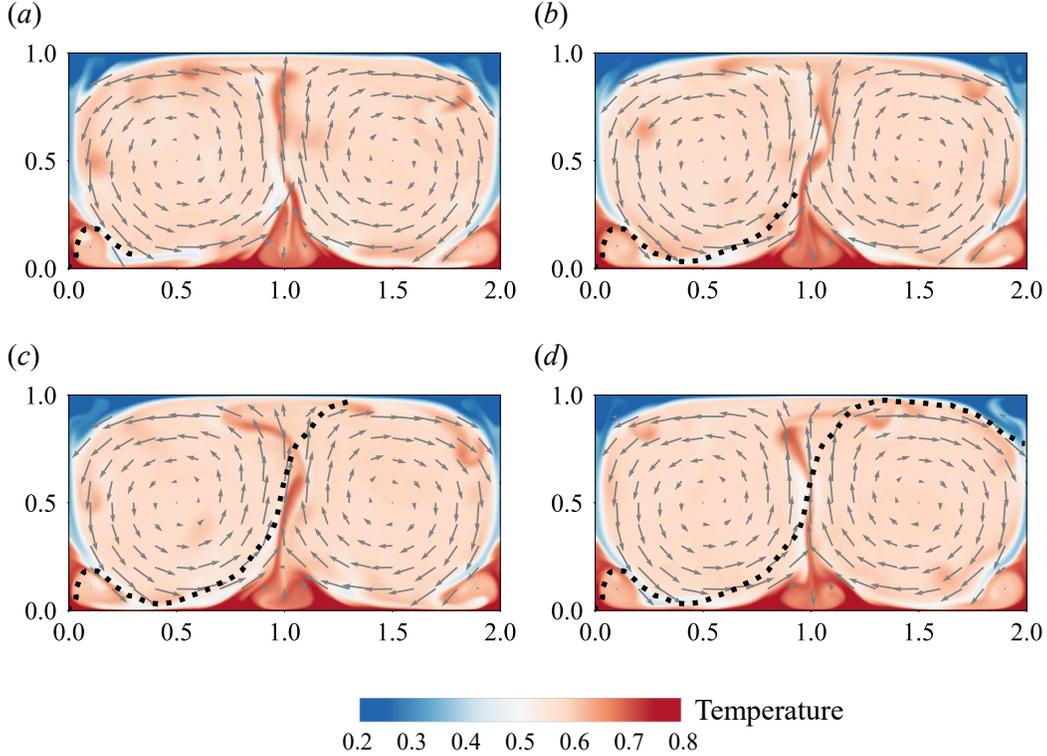}
  \caption{Trajectory (black dotted line) of the smart agent in the RB convection at
  (\textit{a}) $t = 18$, (\textit{b}) $t = 36$, (\textit{c}) $t = 54$, and (\textit{d}) $t = 72$.
  The contour shows the typical instantaneous temperature field, and the vectors denote the velocity field of the convection. \label{fig:instant-traj-Gamma2}}
\end{figure}

We first train the agent to migrate across the turbulent RB cell with $\Gamma=2$.
The agent starts from the point of (0, 0) at the bottom-left corner of the cell, and its goal is to reach the right-side boundary of the cell (i.e., the vertical line of $x = 2$).
Here the simple $\Gamma=2$ cell consists of the characteristic large-scale coherent structure (i.e., a clockwise rotating primary roll and an anticlockwise rotating primary roll), and thus it serves as a paradigm environment for the learning agent.
In Fig. \ref{fig:instant-traj-Gamma2}, we show the instantaneous trajectories of the smart agent after training, and the corresponding video can be viewed in the supplementary movie \cite{SM}.
Initially, the agent moves upward driven by the clockwise rotating corner roll at the bottom-left corner [see Fig. \ref{fig:instant-traj-Gamma2}(a)].
When the agent reaches the edge of the primary roll, which is rotating in the anticlockwise direction, it moves along with the horizontal currents [see Fig. \ref{fig:instant-traj-Gamma2}(b)], until it meets the rising thermals.
The agent then rises on the thermals and ascends higher [see Fig. \ref{fig:instant-traj-Gamma2}(c)].
After reaching the top layer of the cell, due to the right-directed propelling velocity, the agent moves rightward and utilizes the horizontal currents [see Fig. \ref{fig:instant-traj-Gamma2}(d)].
The migration task is completed when the agent reaches the right-side boundary of the cell.
Overall, the smart agent tries to follow the carrier currents as much as possible, and it discovers an effective policy of moving along the edge of the rolls.

\begin{figure}
  \centering
  \includegraphics[width=14cm]{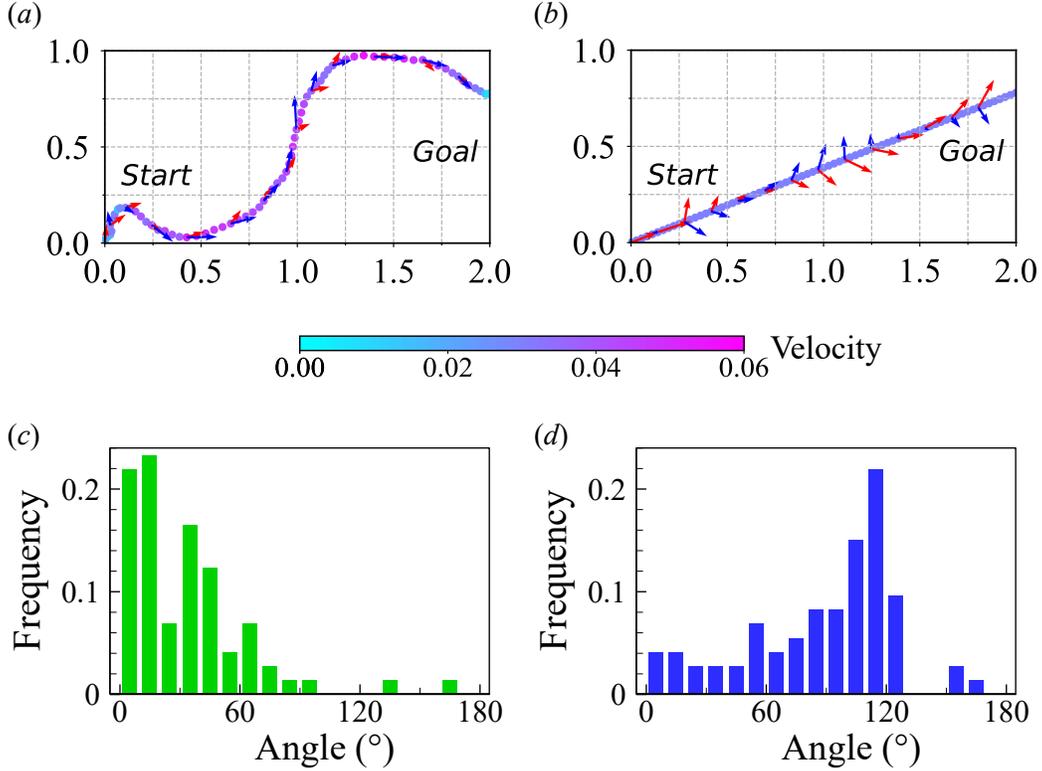}
  \caption{[(\textit{a}) and (\textit{b})] Trajectories for the smart agent enabling energy-efficient migration and
  a naive agent moving straightly, respectively.
  The red arrows denote the propelling velocity and the blue arrows denote the fluid velocity.
  The trajectories are color-coded by the instantaneous velocity magnitude of the agent.
  [(\textit{c}) and (\textit{d})] Histogram of angles between the propelling velocity vector and the fluid velocity vector for
   the smart agent and the naive agent, respectively. \label{fig:compare-traj-Gamma2}}
\end{figure}

A comparison between the smart agent and the naive agent is performed to highlight the differences in propelling behaviors and the savings in energy expenditure.
Here the naive agent refers to the agent that moves straight from the origin to the destination.
We set the naive agent to spend the same amount of total time  $t_{\text{total}}$ as that of the smart agent to complete the migration task,
and its destination point is also the same as the point where the smart agent left the right-side boundary.
Thus, the velocity of the naive agent keeps a constant direction pointing from the origin to the destination, and it keeps a constant magnitude of  $\| \mathbf{u}_{\text{agent}} \|=\| \mathbf{x}_{\text{goal}}-\mathbf{x}_{\text{start}} \|/t_{\text{total}}$.
In Figs. \ref{fig:compare-traj-Gamma2}(a) and \ref{fig:compare-traj-Gamma2}(b), we show trajectories of the smart agent and the naive agent, respectively.
From the instantaneous velocity magnitude shown on the color-coded trajectories, we can see that the naive agent generally migrates slower than the smart agent, because the naive agent travels a shorter distance during the same  $t_{\text{total}}$.
Although the smart agent migrates faster, it does not indicate that the smart agent will consume more energy, since the smart agent can utilize the carrier flow currents to save energy.
Along with the trajectories, we also plot the propelling velocity vector (denoted by the red arrows) and the fluid velocity vector (denoted by the blue arrows).
We calculate the correlation coefficient between the orientation of the propelling velocity vector (i.e., $\theta_{\text{propel}}$) and the orientation of the fluid velocity vector (i.e., $\theta_{\text{fluid}}$) as
\begin{equation}
C=\langle \left[\theta_{\text{propel}}(t)-\langle \theta_{\text{propel}} \rangle  \right]
\left[\theta_{\text{fluid}}(t)-\langle \theta_{\text{fluid}} \rangle  \right] \rangle
/\left( \sigma_{\theta_{\text{propel}}} \sigma_{\theta_{\text{fluid}}} \right).
\end{equation}
Here the orientation is in the range from $-180^{\circ}$ to 180$^{\circ}$.
The positive orientation is defined as anticlockwise rotating the $x$ axis, and the negative orientation is defined as clockwise rotating the $x$ axis.
For the smart agent, the resulting correlation coefficient of 0.56 suggests positive statistical relevance between them, revealing that the smart agent adjusts its migration direction in response to the changing carrier flow, enabling energy-efficient migration;
for the naive agent, the resulting correlation coefficient of $-0.74$ implies negative statistical relevance.
In addition, to quantitatively describe the angles between the propelling velocity vector and the fluid velocity vector, in Figs. \ref{fig:compare-traj-Gamma2}(c) and \ref{fig:compare-traj-Gamma2}(d),
we plot the histogram of those angles for the smart agent and the naive agent, respectively.
We can see for the smart agent, the angles are generally less than 90$^{\circ}$, and the frequency exhibits a peak around 20$^{\circ}$,
which is another evidence that the agent tries to follow the carrier currents.
For the naive agent, the frequency of those angles exhibits a peak around 120$^{\circ}$, indicating that the naive agent has to generate propelling velocity against the carrier flow currents to keep the shortest migrating path.

\begin{figure}
  \centering
  \includegraphics[width=14cm]{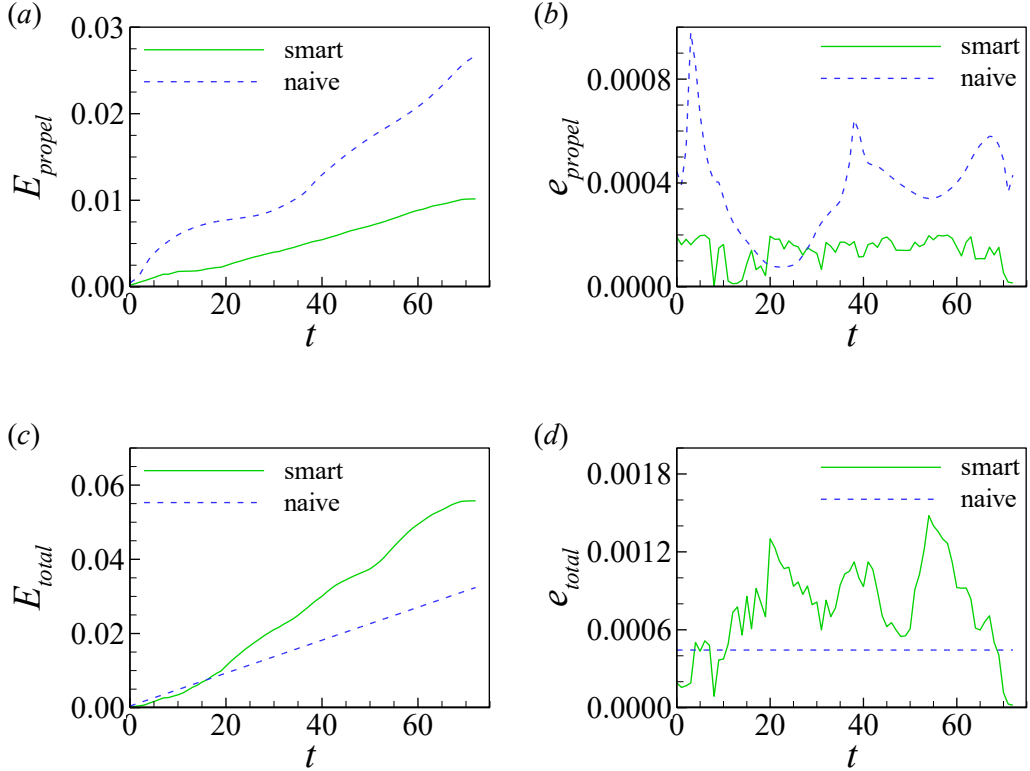}
  \caption{Comparison of the (\textit{a}) accumulative energy consumption $E_{\text{propel}}$,
  (\textit{b}) instantaneous energy consumption $e_{\text{propel}}$,
  (\textit{c}) accumulative total kinetic energy $E_{\text{total}}$,
  and (\textit{d}) instantaneous total kinetic energy $e_{\text{total}}$
  for the smart agent and the naive agent. \label{fig:compare-energy-Gamma2}}
\end{figure}

We assume that only the propelling velocity $\mathbf{u}_{\text{propel}}$ affects the propelling energy consumption for the agents.
Thus, the accumulative energy consumption is calculated as
\begin{equation}
E_{\text{propel}}(t)=\int_{0}^{t}\frac{1}{2} \| \mathbf{u}_{\text{propel}} (\tau) \|^{2} d\tau
\end{equation}
In Fig. \ref{fig:compare-energy-Gamma2}(a), we plot the time series of accumulative energy consumed by the agents.
After completing the migration task, the smart agent consumed around 38\% of the propelling energy compared to that of the naive agent, meaning migrating in the shortest path does not always save energy.
We also calculate the instantaneous energy consumption as
\begin{equation}
e_{\text{propel}}(t)=\frac{1}{2}\| \mathbf{u}_{\text{propel}} (t) \|^{2}
\end{equation}
From Fig. \ref{fig:compare-energy-Gamma2}(b), we can see the $e_{\text{propel}}$  for the smart agent is generally lower than that of the naive agent,
and the $e_{\text{propel}}$  keeps smaller values for the smart agent during the whole migration process.
For the naive agent,  $e_{\text{propel}}$ exhibits a first peak around $t \approx 3$, when it moves across the edges of the left-bottom corner roll that requires a high energy barrier.
At around $t \approx 23$,  $e_{\text{propel}}$ drops to the minimum, because the naive agent reaches the location where flow currents are in alignment with the migration direction.
The second peak of  $e_{\text{propel}}$ appears at around $t \approx 38$, when the naive agent crosses the edge of the primary roll and drifts to the clockwise rotating primary roll.
The third peak of $e_{\text{propel}}$  appears at around $t \approx 67$, when the naive agent approaches the right-side boundary.
We also compare the accumulative total kinetic energy of the agents [see Fig. \ref{fig:compare-energy-Gamma2}(c)], which is calculated as
\begin{equation}
E_{\text{total}}(t)=\int_{0}^{t}\frac{1}{2}\| \mathbf{u}_{\text{agent}} \|^{2} d\tau
\end{equation}
After completing the migration task, the total kinetic energy of the smart agent is almost twice that of the naive agent,
mostly contributed by the kinetic energy of the carrier flow.
Similarly, we calculate the instantaneous total kinetic energy $e_{\text{total}}$ as
\begin{equation}
e_{\text{total}}(t)=\frac{1}{2}\| \mathbf{u}_{\text{agent}} (t) \|^{2}
\end{equation}
From Fig. \ref{fig:compare-energy-Gamma2}(d), we can see the $e_{\text{total}}$ of the smart agent is generally higher than that of the naive agent and keeps larger values during the whole migration process.
Three peaks appear when the smart agents migrate in alignment with the flow direction, thus utilizing more kinetic energy of the carrier flow.
On the other hand, the $e_{\text{total}}$ of the naive agent keeps constant due to the constant value of $\| \mathbf{u}_{\text{agent}} (t) \|$  in the simulation settings.

Choosing appropriate environment cues that the agent can observe is crucial in the RL training.
Here we numerically determine the set of observation variables by comparing the evolution of cumulative reward during the training process.
We train five different instances of each observation variable set with different random seeds, and each set performs one evaluation rollout every 1000 environment steps.
The solid curves correspond to the mean and the shaded region to the minimum and maximum returns over the five trials, and it represents a moving average with a window of 20 time steps.
We first consider the agent is flow-blinded and cannot sense the surrounding flow and temperature information.
The agent can have access to its position information, but only the vertical component, i.e., $s = \{y\}$.
Here we do not consider the horizontal position information (i.e., $x \notin \{s\}$ );
otherwise, the agent trained in the $\Gamma=2$ cell would fail to find optimal trajectory in a larger $\Gamma$ cell once its horizontal position is $x > 2$.
As shown in Fig. \ref{fig:select-observations}(a), the agent performs poorly in the case of flow blinded, which shows similar behavior as that of navigating through unsteady cylinder flow \cite{gunnarson2021learning}.
We then consider the agent can also sense the carrier flow velocity, i.e., $s = \{y, u, v\}$, and plot the results in Fig. \ref{fig:select-observations}(b).
We can see that the agent performs much better, and the cumulative reward is higher than that of the flow-blinded agent.
In addition, we consider the agent can sense extra vorticity information (i.e., $s = \{y, u, v, \omega\}$) \cite{colabrese2017flow,gustavsson2017finding}, and plot the results in Fig. \ref{fig:select-observations}(c).
With the consideration of additional flow field information, the cumulative reward converges at an earlier time (i.e., $t \approx 0.5\times 10^{5}$ for $s = \{y, u, v, \omega \}$)
compared to that of velocity information (i.e., $t \approx 1.0\times 10^{5}$ for $s = \{y, u, v\}$).
On the other hand, in the turbulent RB convection, the temperature acts as an active scalar that influences the velocity, we next consider the agent can sense extra temperature information (i.e., $s = \{ y, u, v, T \}$) \cite{xu2022migration}.
As shown in Fig. \ref{fig:select-observations}(d), the agent outperforms previous ones and the cumulative reward converges at an earlier time and almost remains steady,
suggesting temperature is an important environment cue for the agent to migrate in thermal convection.
In the Appendix \ref{sec:observation}, we evaluate more combinations and then choose $s = \{y, u, v, T\}$ as observation variables.
On the other hand, Kubo and Shimizu \cite{kubo2022efficient} {proposed a framework that can perform fluid flow control with partial observables.
We expect the extension of that framework to turbulent flows will simplify the selection of observables.

\begin{figure}
  \centering
  \includegraphics[width=14cm]{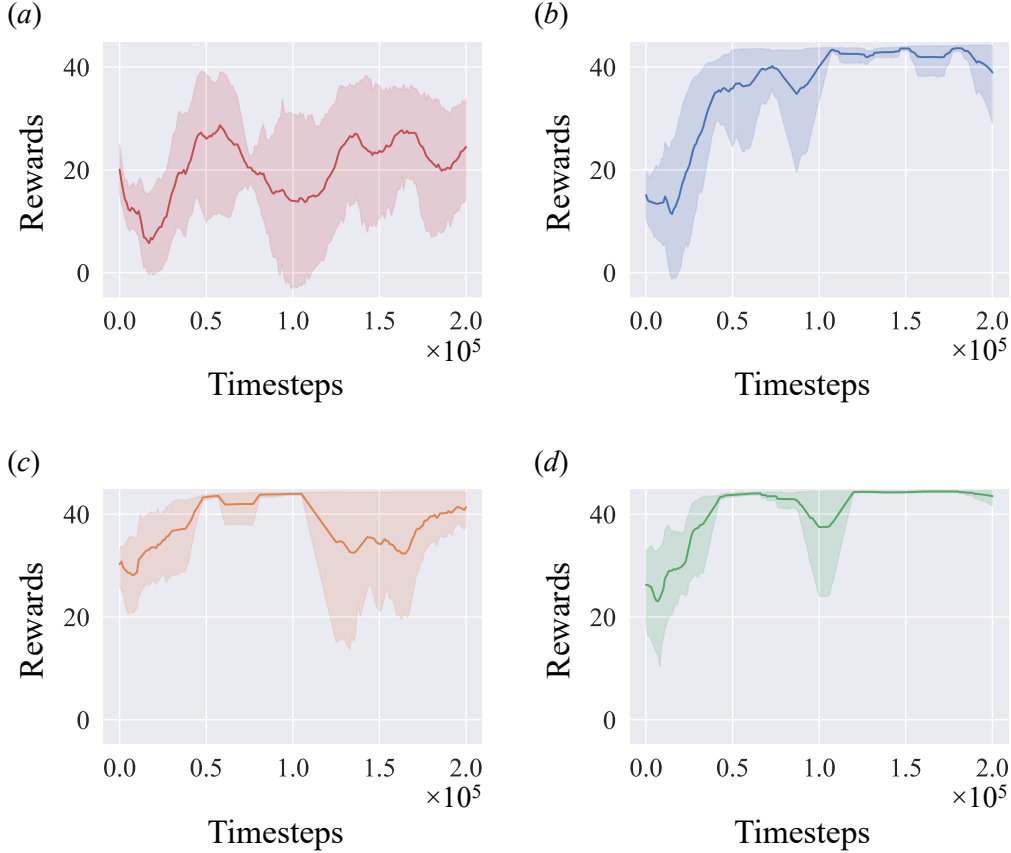}
  \caption{Evolution of the cumulative reward during training for different combinations of observation variables:
  (\textit{a}) $s = \{y\}$, (\textit{b}) $s = \{y, u, v\}$, (\textit{c}) $s = \{y, u, v, \omega \}$, and (\textit{d}) $s = \{y, u, v, T\}$. \label{fig:select-observations}}
\end{figure}

\subsection{Testing the agent to migrate in the RB cell with a larger aspect ratio \label{sec:testing}}
We now apply the obtained policy to test whether the smart agent can migrate in an energy-efficient way in convection cells with larger $\Gamma$.
The flow mode analysis presented in Sec. \ref{sec:flow} indicates that in a larger $\Gamma$ cell,
the dominant flow modes of horizontally stacked rolls are less stable, and the energy contained in higher-order flow modes increases.
Despite these challenges brought by the complex flow features, we can still obtain optimized trajectories, as shown in Fig. \ref{fig:traj_Gamma-4-32}.
Starting from the origin of (0, 0) point, the smart agent first escapes the corner roll at the left-bottom cell corner,
it then ascends higher and rises on the thermal (if the first primary roll is clockwise rotating),
or follows the horizontal currents (if the first primary roll is anticlockwise rotating).
Afterward, the smart agent always tries to migrate along the edges of the primary rolls, where the carrier fluid flows fast and plenty of kinetic energy from the flow is available.
\begin{figure}
  \centering
  \includegraphics[width=16cm]{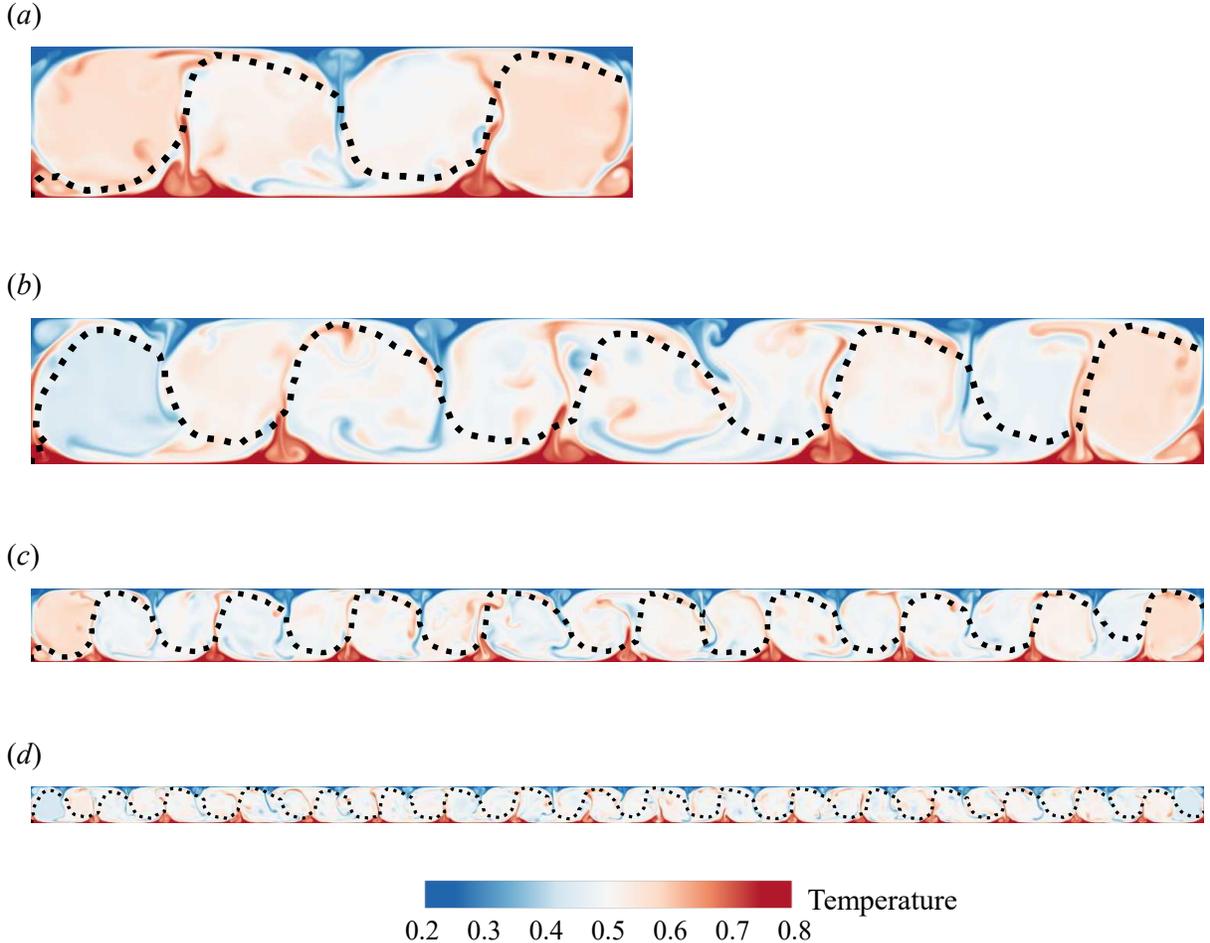}
  \caption{Trajectory (black dotted line) of the smart agent in the convection cell with
  (\textit{a}) $\Gamma=4$, (\textit{b}) $\Gamma=8$, (\textit{c}) $\Gamma=16$, and (\textit{d}) $\Gamma=32$.
  The contour shows the typical instantaneous temperature field. \label{fig:traj_Gamma-4-32}}
\end{figure}

We then compare the propelling energy consumed by the smart agent and the naive agent in the convection cell with various $\Gamma$.
In Fig. \ref{fig:energy-Gamma-4-32}(a), we plot the accumulative propelling energy for the agents when they complete the migration task.
Generally, for both agents, the energy consumption increases with the increase of $\Gamma$, due to longer migration distance.
For the smart agent, it enables an energy-efficient migration strategy via migrating along the edges of horizontally stacked multiple primary rolls, thus we have $E_{\text{propel}} \propto \Gamma$.
To quantitatively describe how much propelling energy can be saved, we plot the ratio of energy consumed by the smart agent to that of the naive agent, as shown in Fig. \ref{fig:energy-Gamma-4-32}(b).
We can see that in a larger $\Gamma$ cell, the ratio of  $E_{\text{smart}}/E_{\text{naive}}$ is smaller,
meaning more propelling energy can be saved by the smart agent.
The reason is that in a larger $\Gamma$ cell, the naive agent has to cross more edges of the circulation rolls and overcome higher energy barriers,
while the smart agent follows the carrier currents in an energy-efficient way.
\begin{figure}
  \centering
  \includegraphics[width=14cm]{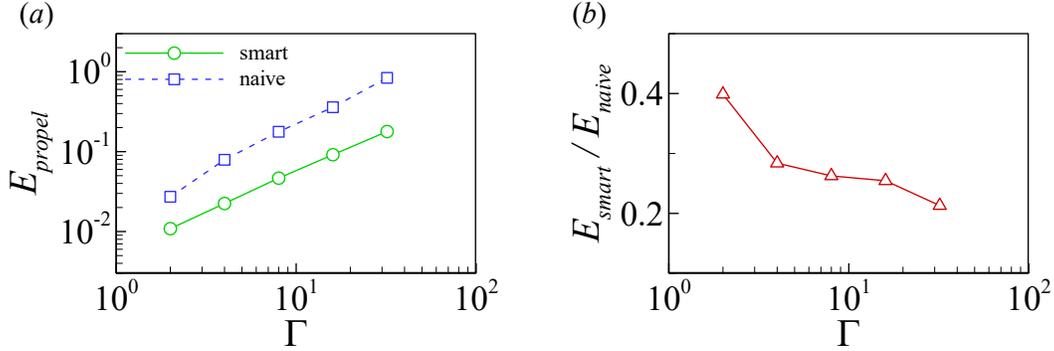}
  \caption{(\textit{a}) The propelling energy consumed by the smart agent and the naive agent,
  (\textit{b}) the ratio of energy consumed by the smart agent to that of the naive agent as a function of $\Gamma$. \label{fig:energy-Gamma-4-32}}
\end{figure}

The above results are obtained with the prescribed and fixed origin position, namely, the location at the (0, 0) point.
We further test the robustness of the energy-efficient policy concerning random origin position.
We first release the agent in the position where the local flow velocity is weak, i.e., $\| \mathbf{u}_{\text{fluid}} \| < 0.03$, and 100 example trajectories are plotted in Fig. \ref{fig:success-fail-traj}.
We can see regardless of the random origin position, the successful attempts gradually converge to a similar path line, and the agent migrates along the edges of the primary rolls.
It should be noted that we restrict these "random" positions to be $x < \Gamma/2$,
which prevents the agent from being released too close to the outlet.
The average success rate to complete the migration task in the $\Gamma = 4,8,16$, and 32 cell is 74\%, 87\%, 92\%, and 97\%, respectively.
We then release the agent in the position where the local flow velocity is strong, i.e., $\| \mathbf{u}_{\text{fluid}} \| > 0.03$.
The average success rate is much higher, and it is 100\%, 100\%, 100\%, and 99\% in the $\Gamma=4,8,16$, and 32 cell, respectively.
For the sake of clarity, we do not repeat plotting the example trajectories.
\begin{figure}
  \centering
  \includegraphics[width=16cm]{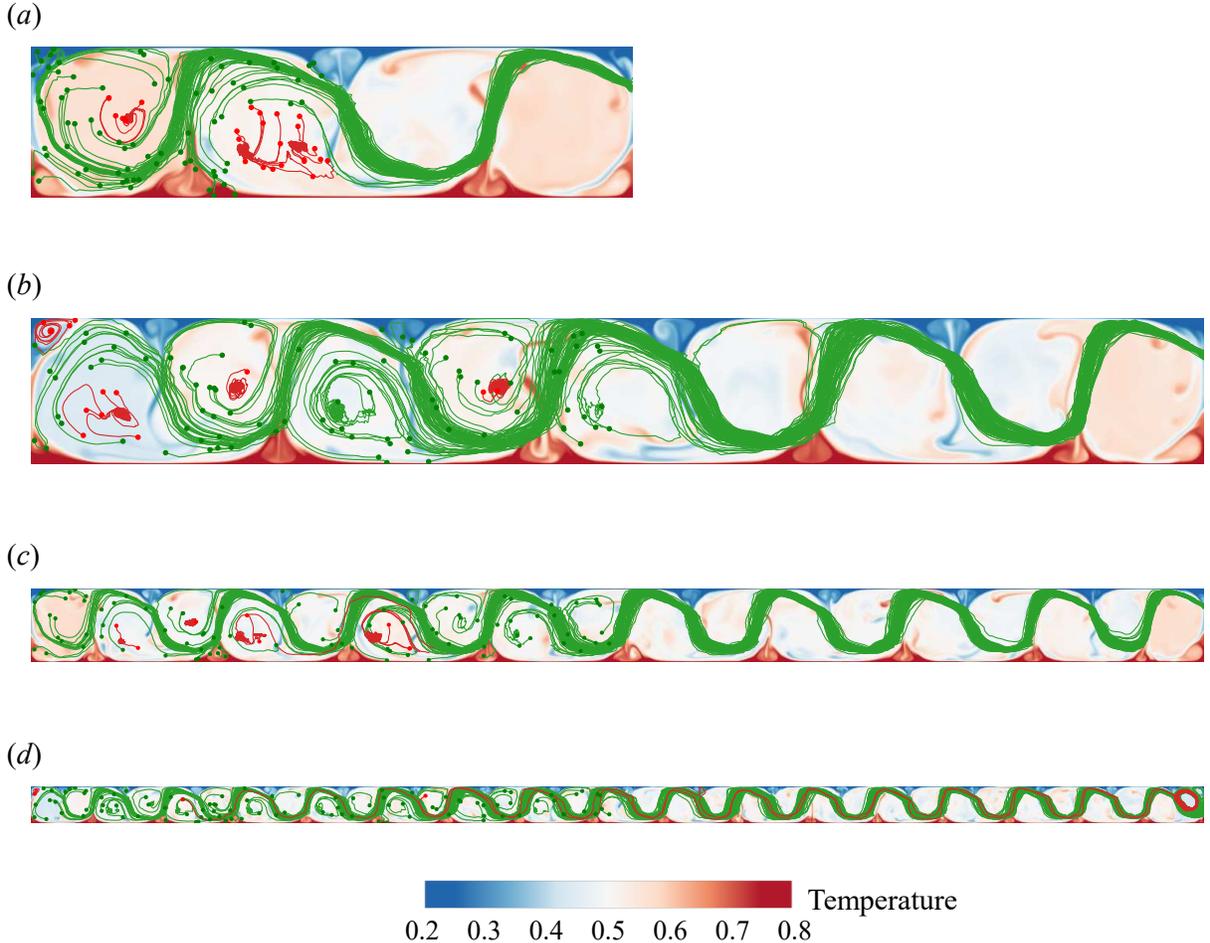}
  \caption{Example trajectories in the
  (\textit{a}) $\Gamma=4$, (\textit{b}) $\Gamma=8$, (\textit{c}) $\Gamma=16$, and (\textit{d}) $\Gamma=32$ cell (the agents are released where $\|\mathbf{u}_{\text{fluid}} \|<0.03$).
  Green lines represent successful attempts to complete the migration, while red lines represent unsuccessful attempts.
  The contour shows the typical instantaneous temperature field. \label{fig:success-fail-traj}}
\end{figure}

The above results indicate that the success rate to complete the migration task increases with the increase of $\Gamma$.
On the other hand, in a larger $\Gamma$ cell, the flow structure is more complex, and we expect it would be more challenging for the agent to complete the migration task and earn a higher success rate.
To understand such behaviors, we further consider the following scenarios:
(i) the agents being released where carrier flow velocity is $\| \mathbf{u}_{\text{fluid}} \| < 0.01$,
(ii) the agents being released where carrier flow velocity is $0.01 < \| \mathbf{u}_{\text{fluid}} \| < 0.02$,
and (iii) the agents being released where carrier flow velocity is $0.02 < \| \mathbf{u}_{\text{fluid}} \| < 0.03$.
We can see from Fig. \ref{fig:success-rate} that in convection cells with the same $\Gamma$, the average success rate is higher if the carrier flow velocity at the origin is stronger.
Because the global flow strength is stronger in a larger $\Gamma$ cell (see discussion in Sec. \ref{sec:flow}), the agents are more likely to be released where carrier flow is strong.
Utilizing stronger carrier flow currents, the agents can complete the migration task within a shorter time and earn a higher reward, thus the agent is more likely to repeat that action in the future.
The higher sampling frequency for the agent in areas with stronger flow strength leads to higher success rate in larger $\Gamma$ cell.
It should be noted that such a trend is only obvious when the origin of agents possesses weak carrier flow velocity;
with strong carrier flow velocity, the success rates are always near 100\%.

\begin{figure}
  \centering
  \includegraphics[width=9cm]{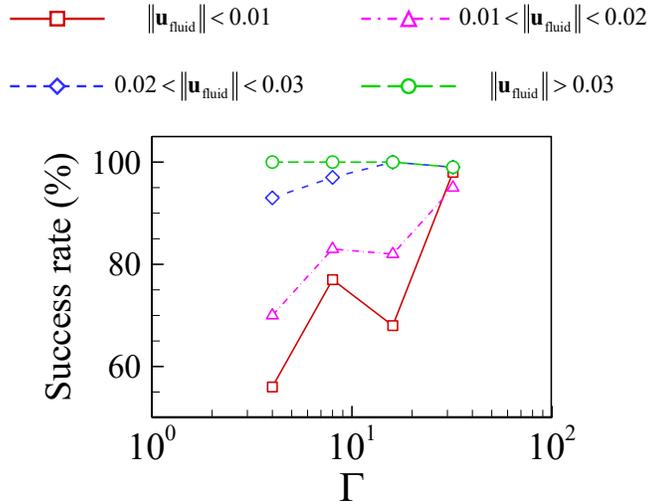}
  \caption{Average success rate as functions of $\Gamma$ when the agents are randomly released at the positions with the different local carrier flow velocities. \label{fig:success-rate}}
\end{figure}

\section{Conclusions \label{sec:conclusions}}
In this work, using the reinforcement learning algorithm, we performed numerical training of the self-propelling agent migrating long distance in a thermal turbulent environment.
We choose the paradigmatic turbulent RB convection cell as the flow environment, which can incorporate strong fluctuations of velocity and temperature.
To build up the reinforcement learning framework, we designed a reward function that simultaneously considers the current state, energy consumption, and time consumption of the agent.
We also compare the evolution of cumulative reward for different combinations of observation variables.
We select the position of the agent, as well as the velocity and temperature of the carrier flow as appropriate environmental cues.
The simulation results in a $\Gamma = 2$ RB cell showed that, compared to a naive agent that moves straight from the origin to the destination,
the smart agent can learn to utilize the carrier flow currents to save propelling energy.

We then apply the optimal policy obtained in the $\Gamma = 2$ cell and test the smart agent migrating in convection cells with larger $\Gamma$.
From flow mode analysis, we found the dominant flow modes in a larger $\Gamma$ RB cell consist of less stable horizontally stacked rolls, and the energy contained in higher-order flow modes increases with the increase of $\Gamma$.
Although these complex flow features bring challenges to optimizing the trajectories for the smart agent, we can still obtain energy-efficient migrating trajectories using the policy trained in the $\Gamma = 2$ RB cell.
In addition, we found the ratio of propelling energy consumed by the smart agent to that of the naive agent decreases with the increase of $\Gamma$,
meaning more propelling energy can be saved by the smart agent in a larger $\Gamma$ cell.
The reason is that in a larger $\Gamma$ cell, the naive agent has to cross more edges of the circulation rolls and overcome higher energy barriers, while the smart agent always tries to follow the carrier currents as much as possible.

We also evaluate the optimized policy when the agents are being released from the randomly chosen origin, which aims to test the robustness of the learning framework.
We found the success rate increases with the increase of $\Gamma$, despite the flow structures being more complex in a larger $\Gamma$ cell.
The main reason is that, in a larger $\Gamma$ cell, the global flow strength is stronger (evident by the relationship between Re and $\Gamma$),
and the agent is more likely to be released in positions where the carrier flow velocity is stronger.
Utilizing stronger carrier flow velocity, the agent can complete the migration task within a shorter time and receive a higher reward,
thus leading to a higher success rate.
Our work has implications for long-distance migration problems, for example, the UAVs patrolling in the convective layer of the atmosphere.
Migrating in energy-efficient trajectories, the UAVs can increase endurance and cover a wider range.

\begin{acknowledgments}
This work was supported by the National Natural Science Foundation of China (NSFC) through Grants No. 12272311 and No. 12125204,
the National Key Project via No. GJXM92579,
and the 111 project of China (Project No. B17037).
The authors acknowledge the Beijing Beilong Super Cloud Computing Co., Ltd for providing HPC resources that have contributed to the research results reported within this paper.
\end{acknowledgments}

\appendix

\section{Sensitivity of the hyperparameters in the reward function \label{sec:sensitivity}}

In our designed reward function [see Eqs. (\ref{Eq.reward})-(\ref{Eq.reward-rh})], we have two hyperparameters:
one is  $\phi$, which represents the penalty when the agent is out of the flow domain;
the other is  $\varepsilon$, which is the reward scale coefficient.
We tuned these two parameters separately to determine the optimal hyperparameters.
It should be noted that we compared the value of the normalized reward (i.e., varied between 0 and 1) rather than the absolute value of the reward.
We can see from Fig. \ref{fig:param_dependence}(a) that, small values of the penalty $\phi$ (e.g., $\phi =$ 1 and 5) results in substantial degradation of performance;
large values of the penalty (e.g.,  $\phi \ge$ 10) almost lead to the same performance.
As for the reward scale coefficient  $\varepsilon$, it is almost insensitive for the investigated value and they can all give optimal policy, as shown in Fig. \ref{fig:param_dependence}(b).
However, a large value (e.g., $\varepsilon=100$, not shown here for clarity) would result in  $\sum (r_{s}+r_{e}) > \sum r_{h}$ during training, and the agent's failure to explore the successful trajectory within the given time of  $t_{\max}$.
\begin{figure}
 \centering
 \includegraphics[width=14cm]{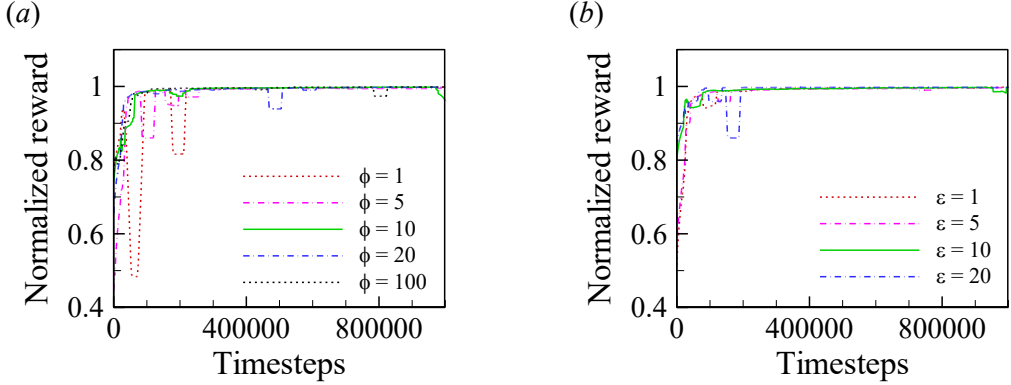}
 \caption{Sensitivity of the hyperparameters in the reward function:
 (\textit{a}) the penalty $\phi$ when the agent is out of the flow domain and (\textit{b}) the reward scale coefficient  $\varepsilon$.
 \label{fig:param_dependence}}
\end{figure}

\section{Evaluation of different combinations of observation variables\label{sec:observation}}
In addition to the observation variables described in Sec. \ref{sec:traning},
we also consider the following different combinations:
(i) The agent has access to position and velocity information, and it can sense strain rate \cite{qiu2022navigation,qiu2022active}, i.e., $s=\{ y, u, v, s_{xx} \}$ and $s=\{ y, u, v, s_{xy} \}$.
Here $s_{xx}=\partial_{x}u$ and $s_{xy}=(\partial_{y}u+\partial_{x}v)/2$.
We did not consider the $s_{yy}=\partial_{y}v$ component of the strain rate tensor, because flow continuity equation gives $s_{xx}+s_{yy}=0$ in 2D flows, which means $s_{xx}$ and $s_{yy}$ are negatively correlated.
(ii) The agent has access to position and velocity information, and it can sense temperature gradient, i.e., $s=\{y, u, v, (\nabla T)_{x}  \}$ and $s=\{y, u, v, (\nabla T)_{y}  \}$.
Here $(\nabla T)_{x}$ essentially represents the vorticity produced by buoyancy in the 2D convection flow \cite{xu2022production}.
(iii) The agent has access to position, velocity and temperature information, and it can sense additional vorticity, strain rate, or temperature gradient,
i.e., $s=\{y, u, v, T, s_{xx} \}$, $s=\{y, u, v, T, s_{xy} \}$, $s=\{y, u, v, T, \omega \}$, $s=\{y, u, v, T, (\nabla T)_{x}\}$, and $s=\{y, u, v, T, (\nabla T)_{y}\}$.
In Fig. \ref{fig:more-observations}, we plot the evolution of the cumulative reward during training for the above nine combinations of observation variables.
We can see these combinations only slightly changes the converging speed of the training, not the asymptotic accumulative reward value.
Among them, the $s=\{y, u, v, T, (\nabla T)_{x}\}$ shown in Fig. \ref{fig:more-observations}(h) outperforms other combinations.
In practical applications, velocity or temperature sensing could be implemented via a variety of methods, such as pitot tubes, hot wire, and so on; while vorticity, shear strain component, and temperature gradient should be computed from several velocities or temperature sensors, which increases the complexity that the agent has to sense.
Thus, as described in Sec. \ref{sec:traning}, we deliberately keep simple the environmental cues of local information $s=\{ y, u, v, T \}$ that the agent can see to guide its migration, such that the amount of data storage by the agent can be reduced in practical applications.

\begin{figure}
 \centering
 \includegraphics[width=16cm]{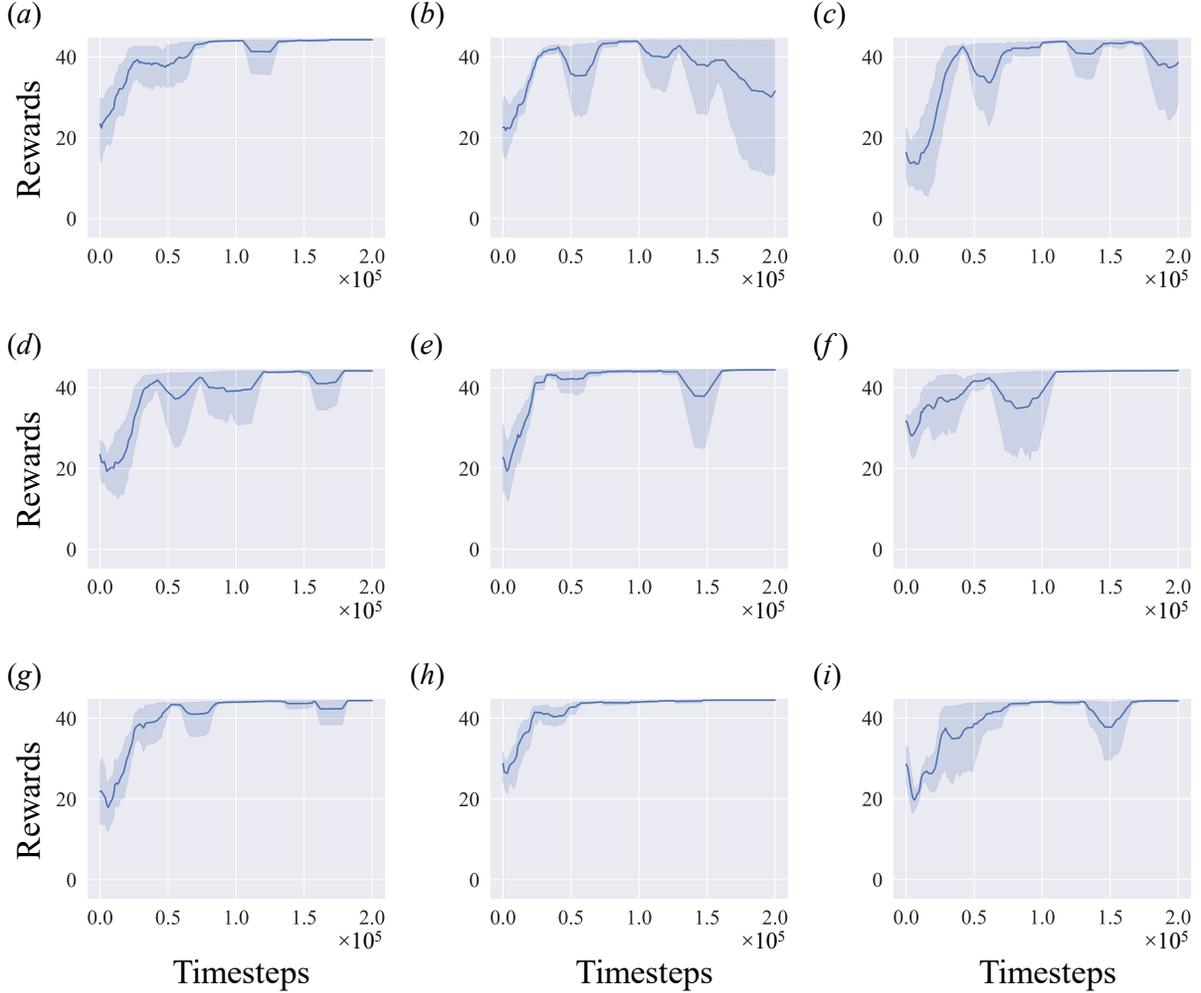}
 \caption{Evolution of the cumulative reward during training for different combinations of observation variables:
 (\textit{a}) $s = \{y, u, v, s_{xx} \}$,
 (\textit{b}) $s = \{y, u, v, s_{xy} \}$,
 (\textit{c}) $s = \{y, u, v, (\nabla T)_{x} \}$,
 (\textit{d}) $s = \{y, u, v, (\nabla T)_{y} \}$,
 (\textit{e}) $s = \{y, u, v, T, s_{xx} \}$,
 (\textit{f}) $s = \{y, u, v, T, s_{xy} \}$,
 (\textit{g}) $s = \{y, u, v, T, \omega \}$,
 (\textit{h}) $s = \{y, u, v, T, (\nabla T)_{x}\}$,
 and (\textit{i}) $s = \{y, u, v, T, (\nabla T)_{y}\}$. \label{fig:more-observations}}
\end{figure}


\begin{thebibliography}{61}%
\makeatletter
\providecommand \@ifxundefined [1]{%
 \@ifx{#1\undefined}
}%
\providecommand \@ifnum [1]{%
 \ifnum #1\expandafter \@firstoftwo
 \else \expandafter \@secondoftwo
 \fi
}%
\providecommand \@ifx [1]{%
 \ifx #1\expandafter \@firstoftwo
 \else \expandafter \@secondoftwo
 \fi
}%
\providecommand \natexlab [1]{#1}%
\providecommand \enquote  [1]{``#1''}%
\providecommand \bibnamefont  [1]{#1}%
\providecommand \bibfnamefont [1]{#1}%
\providecommand \citenamefont [1]{#1}%
\providecommand \href@noop [0]{\@secondoftwo}%
\providecommand \href [0]{\begingroup \@sanitize@url \@href}%
\providecommand \@href[1]{\@@startlink{#1}\@@href}%
\providecommand \@@href[1]{\endgroup#1\@@endlink}%
\providecommand \@sanitize@url [0]{\catcode `\\12\catcode `\$12\catcode
  `\&12\catcode `\#12\catcode `\^12\catcode `\_12\catcode `\%12\relax}%
\providecommand \@@startlink[1]{}%
\providecommand \@@endlink[0]{}%
\providecommand \url  [0]{\begingroup\@sanitize@url \@url }%
\providecommand \@url [1]{\endgroup\@href {#1}{\urlprefix }}%
\providecommand \urlprefix  [0]{URL }%
\providecommand \Eprint [0]{\href }%
\providecommand \doibase [0]{https://doi.org/}%
\providecommand \selectlanguage [0]{\@gobble}%
\providecommand \bibinfo  [0]{\@secondoftwo}%
\providecommand \bibfield  [0]{\@secondoftwo}%
\providecommand \translation [1]{[#1]}%
\providecommand \BibitemOpen [0]{}%
\providecommand \bibitemStop [0]{}%
\providecommand \bibitemNoStop [0]{.\EOS\space}%
\providecommand \EOS [0]{\spacefactor3000\relax}%
\providecommand \BibitemShut  [1]{\csname bibitem#1\endcsname}%
\let\auto@bib@innerbib\@empty
\bibitem [{\citenamefont {Weimerskirch}\ \emph {et~al.}(2016)\citenamefont
  {Weimerskirch}, \citenamefont {Bishop}, \citenamefont {Jeanniard-du Dot},
  \citenamefont {Prudor},\ and\ \citenamefont
  {Sachs}}]{weimerskirch2016frigate}%
  \BibitemOpen
  \bibfield  {author} {\bibinfo {author} {\bibfnamefont {H.}~\bibnamefont
  {Weimerskirch}}, \bibinfo {author} {\bibfnamefont {C.}~\bibnamefont
  {Bishop}}, \bibinfo {author} {\bibfnamefont {T.}~\bibnamefont {Jeanniard-du
  Dot}}, \bibinfo {author} {\bibfnamefont {A.}~\bibnamefont {Prudor}},\ and\
  \bibinfo {author} {\bibfnamefont {G.}~\bibnamefont {Sachs}},\ }\bibfield
  {title} {\bibinfo {title} {Frigate birds track atmospheric conditions over
  months-long transoceanic flights},\ }\href@noop {} {\bibfield  {journal}
  {\bibinfo  {journal} {Science}\ }\textbf {\bibinfo {volume} {353}},\ \bibinfo
  {pages} {74} (\bibinfo {year} {2016})}\BibitemShut {NoStop}%
\bibitem [{\citenamefont {Williams}\ \emph {et~al.}(2020)\citenamefont
  {Williams}, \citenamefont {Shepard}, \citenamefont {Holton}, \citenamefont
  {Alarc{\'o}n}, \citenamefont {Wilson},\ and\ \citenamefont
  {Lambertucci}}]{williams2020physical}%
  \BibitemOpen
  \bibfield  {author} {\bibinfo {author} {\bibfnamefont {H.~J.}\ \bibnamefont
  {Williams}}, \bibinfo {author} {\bibfnamefont {E.}~\bibnamefont {Shepard}},
  \bibinfo {author} {\bibfnamefont {M.~D.}\ \bibnamefont {Holton}}, \bibinfo
  {author} {\bibfnamefont {P.}~\bibnamefont {Alarc{\'o}n}}, \bibinfo {author}
  {\bibfnamefont {R.}~\bibnamefont {Wilson}},\ and\ \bibinfo {author}
  {\bibfnamefont {S.}~\bibnamefont {Lambertucci}},\ }\bibfield  {title}
  {\bibinfo {title} {Physical limits of flight performance in the heaviest
  soaring bird},\ }\href@noop {} {\bibfield  {journal} {\bibinfo  {journal}
  {Proc. Natl. Acad. Sci. U.S.A.}\ }\textbf {\bibinfo {volume} {117}},\
  \bibinfo {pages} {17884} (\bibinfo {year} {2020})}\BibitemShut {NoStop}%
\bibitem [{\citenamefont {Croxall}\ \emph {et~al.}(2005)\citenamefont
  {Croxall}, \citenamefont {Silk}, \citenamefont {Phillips}, \citenamefont
  {Afanasyev},\ and\ \citenamefont {Briggs}}]{croxall2005global}%
  \BibitemOpen
  \bibfield  {author} {\bibinfo {author} {\bibfnamefont {J.~P.}\ \bibnamefont
  {Croxall}}, \bibinfo {author} {\bibfnamefont {J.~R.}\ \bibnamefont {Silk}},
  \bibinfo {author} {\bibfnamefont {R.~A.}\ \bibnamefont {Phillips}}, \bibinfo
  {author} {\bibfnamefont {V.}~\bibnamefont {Afanasyev}},\ and\ \bibinfo
  {author} {\bibfnamefont {D.~R.}\ \bibnamefont {Briggs}},\ }\bibfield  {title}
  {\bibinfo {title} {Global circumnavigations: tracking year-round ranges of
  nonbreeding albatrosses},\ }\href@noop {} {\bibfield  {journal} {\bibinfo
  {journal} {Science}\ }\textbf {\bibinfo {volume} {307}},\ \bibinfo {pages}
  {249} (\bibinfo {year} {2005})}\BibitemShut {NoStop}%
\bibitem [{\citenamefont {Lancaster}(1885)}]{lancaster1885problem}%
  \BibitemOpen
  \bibfield  {author} {\bibinfo {author} {\bibfnamefont {I.}~\bibnamefont
  {Lancaster}},\ }\bibfield  {title} {\bibinfo {title} {The problem of the
  soaring bird},\ }\href@noop {} {\bibfield  {journal} {\bibinfo  {journal}
  {Am. Nat.}\ }\textbf {\bibinfo {volume} {19}},\ \bibinfo {pages} {1055}
  (\bibinfo {year} {1885})}\BibitemShut {NoStop}%
\bibitem [{\citenamefont {MacCready}(1958)}]{maccready1958optimum}%
  \BibitemOpen
  \bibfield  {author} {\bibinfo {author} {\bibfnamefont {P.~B.}\ \bibnamefont
  {MacCready}},\ }\bibfield  {title} {\bibinfo {title} {Optimum airspeed
  selector},\ }\href@noop {} {\bibfield  {journal} {\bibinfo  {journal}
  {Soaring}\ }\textbf {\bibinfo {volume} {10}},\ \bibinfo {pages} {10}
  (\bibinfo {year} {1958})}\BibitemShut {NoStop}%
\bibitem [{\citenamefont {Allen}(2005)}]{allen2005autonomous}%
  \BibitemOpen
  \bibfield  {author} {\bibinfo {author} {\bibfnamefont {M.}~\bibnamefont
  {Allen}},\ }\bibfield  {title} {\bibinfo {title} {Autonomous soaring for
  improved endurance of a small uninhabitated air vehicle},\ }in\ \href@noop {}
  {\emph {\bibinfo {booktitle} {Proceedings of the 43rd AIAA Aerospace Sciences
  Meeting and Exhibit}}}\ (\bibinfo {year} {AIAA, Reston, VA, 2005})\ p.\
  \bibinfo {pages} {1025}\BibitemShut {NoStop}%
\bibitem [{\citenamefont {Bencatel}\ \emph {et~al.}(2013)\citenamefont
  {Bencatel}, \citenamefont {de~Sousa},\ and\ \citenamefont
  {Girard}}]{bencatel2013atmospheric}%
  \BibitemOpen
  \bibfield  {author} {\bibinfo {author} {\bibfnamefont {R.}~\bibnamefont
  {Bencatel}}, \bibinfo {author} {\bibfnamefont {J.~T.}\ \bibnamefont
  {de~Sousa}},\ and\ \bibinfo {author} {\bibfnamefont {A.}~\bibnamefont
  {Girard}},\ }\bibfield  {title} {\bibinfo {title} {Atmospheric flow field
  models applicable for aircraft endurance extension},\ }\href@noop {}
  {\bibfield  {journal} {\bibinfo  {journal} {Prog. Aeosp. Sci.}\ }\textbf
  {\bibinfo {volume} {61}},\ \bibinfo {pages} {1} (\bibinfo {year}
  {2013})}\BibitemShut {NoStop}%
\bibitem [{\citenamefont {Allen}(2006)}]{allen2006updraft}%
  \BibitemOpen
  \bibfield  {author} {\bibinfo {author} {\bibfnamefont {M.}~\bibnamefont
  {Allen}},\ }\bibfield  {title} {\bibinfo {title} {Updraft model for
  development of autonomous soaring uninhabited air vehicles},\ }in\ \href@noop
  {} {\emph {\bibinfo {booktitle} {Proceedings of the 44th AIAA Aerospace
  Sciences Meeting and Exhibit}}}\ (\bibinfo {year} {AIAA, Reston, VA, 2006})\
  p.\ \bibinfo {pages} {1510}\BibitemShut {NoStop}%
\bibitem [{\citenamefont {Lawrance}\ and\ \citenamefont
  {Sukkarieh}(2009)}]{lawrance2009wind}%
  \BibitemOpen
  \bibfield  {author} {\bibinfo {author} {\bibfnamefont {N.}~\bibnamefont
  {Lawrance}}\ and\ \bibinfo {author} {\bibfnamefont {S.}~\bibnamefont
  {Sukkarieh}},\ }\bibfield  {title} {\bibinfo {title} {Wind energy based path
  planning for a small gliding unmanned aerial vehicle},\ }in\ \href@noop {}
  {\emph {\bibinfo {booktitle} {Proceedings of the AIAA Guidance, Navigation,
  and Control Conference}}}\ (\bibinfo {year} {AIAA, Reston, VA, 2009})\ p.\
  \bibinfo {pages} {6112}\BibitemShut {NoStop}%
\bibitem [{\citenamefont {{\'A}kos}\ \emph {et~al.}(2010)\citenamefont
  {{\'A}kos}, \citenamefont {Nagy}, \citenamefont {Leven},\ and\ \citenamefont
  {Vicsek}}]{akos2010thermal}%
  \BibitemOpen
  \bibfield  {author} {\bibinfo {author} {\bibfnamefont {Z.}~\bibnamefont
  {{\'A}kos}}, \bibinfo {author} {\bibfnamefont {M.}~\bibnamefont {Nagy}},
  \bibinfo {author} {\bibfnamefont {S.}~\bibnamefont {Leven}},\ and\ \bibinfo
  {author} {\bibfnamefont {T.}~\bibnamefont {Vicsek}},\ }\bibfield  {title}
  {\bibinfo {title} {Thermal soaring flight of birds and unmanned aerial
  vehicles},\ }\href@noop {} {\bibfield  {journal} {\bibinfo  {journal}
  {Bioinspir. Biomim.}\ }\textbf {\bibinfo {volume} {5}},\ \bibinfo {pages}
  {045003} (\bibinfo {year} {2010})}\BibitemShut {NoStop}%
\bibitem [{\citenamefont {Laurent}\ \emph {et~al.}(2021)\citenamefont
  {Laurent}, \citenamefont {Fogg}, \citenamefont {Ginsburg}, \citenamefont
  {Halverson}, \citenamefont {Lanzone}, \citenamefont {Miller}, \citenamefont
  {Winkler},\ and\ \citenamefont {Bewley}}]{laurent2021turbulence}%
  \BibitemOpen
  \bibfield  {author} {\bibinfo {author} {\bibfnamefont {K.~M.}\ \bibnamefont
  {Laurent}}, \bibinfo {author} {\bibfnamefont {B.}~\bibnamefont {Fogg}},
  \bibinfo {author} {\bibfnamefont {T.}~\bibnamefont {Ginsburg}}, \bibinfo
  {author} {\bibfnamefont {C.}~\bibnamefont {Halverson}}, \bibinfo {author}
  {\bibfnamefont {M.~J.}\ \bibnamefont {Lanzone}}, \bibinfo {author}
  {\bibfnamefont {T.~A.}\ \bibnamefont {Miller}}, \bibinfo {author}
  {\bibfnamefont {D.~W.}\ \bibnamefont {Winkler}},\ and\ \bibinfo {author}
  {\bibfnamefont {G.~P.}\ \bibnamefont {Bewley}},\ }\bibfield  {title}
  {\bibinfo {title} {Turbulence explains the accelerations of an eagle in
  natural flight},\ }\href@noop {} {\bibfield  {journal} {\bibinfo  {journal}
  {Proc. Natl. Acad. Sci. U.S.A.}\ }\textbf {\bibinfo {volume} {118}},\
  \bibinfo {pages} {e2102588118} (\bibinfo {year} {2021})}\BibitemShut
  {NoStop}%
\bibitem [{\citenamefont {Ahlers}\ \emph {et~al.}(2009)\citenamefont {Ahlers},
  \citenamefont {Grossmann},\ and\ \citenamefont {Lohse}}]{ahlers2009heat}%
  \BibitemOpen
  \bibfield  {author} {\bibinfo {author} {\bibfnamefont {G.}~\bibnamefont
  {Ahlers}}, \bibinfo {author} {\bibfnamefont {S.}~\bibnamefont {Grossmann}},\
  and\ \bibinfo {author} {\bibfnamefont {D.}~\bibnamefont {Lohse}},\ }\bibfield
   {title} {\bibinfo {title} {Heat transfer and large scale dynamics in
  turbulent {R}ayleigh-{B}{\'e}nard convection},\ }\href@noop {} {\bibfield
  {journal} {\bibinfo  {journal} {Rev. Mod. Phys.}\ }\textbf {\bibinfo {volume}
  {81}},\ \bibinfo {pages} {503} (\bibinfo {year} {2009})}\BibitemShut
  {NoStop}%
\bibitem [{\citenamefont {Chill{\`a}}\ and\ \citenamefont
  {Schumacher}(2012)}]{chilla2012new}%
  \BibitemOpen
  \bibfield  {author} {\bibinfo {author} {\bibfnamefont {F.}~\bibnamefont
  {Chill{\`a}}}\ and\ \bibinfo {author} {\bibfnamefont {J.}~\bibnamefont
  {Schumacher}},\ }\bibfield  {title} {\bibinfo {title} {New perspectives in
  turbulent {R}ayleigh-{B}{\'e}nard convection},\ }\href@noop {} {\bibfield
  {journal} {\bibinfo  {journal} {Eur. Phys. J. E}\ }\textbf {\bibinfo {volume}
  {35}},\ \bibinfo {pages} {1} (\bibinfo {year} {2012})}\BibitemShut {NoStop}%
\bibitem [{\citenamefont {Xia}(2013)}]{xia2013current}%
  \BibitemOpen
  \bibfield  {author} {\bibinfo {author} {\bibfnamefont {K.-Q.}\ \bibnamefont
  {Xia}},\ }\bibfield  {title} {\bibinfo {title} {Current trends and future
  directions in turbulent thermal convection},\ }\href@noop {} {\bibfield
  {journal} {\bibinfo  {journal} {Theor. Appl. Mech. Lett.}\ }\textbf {\bibinfo
  {volume} {3}},\ \bibinfo {pages} {052001} (\bibinfo {year}
  {2013})}\BibitemShut {NoStop}%
\bibitem [{\citenamefont {Atkinson}\ and\ \citenamefont
  {Wu~Zhang}(1996)}]{atkinson1996mesoscale}%
  \BibitemOpen
  \bibfield  {author} {\bibinfo {author} {\bibfnamefont {B.}~\bibnamefont
  {Atkinson}}\ and\ \bibinfo {author} {\bibfnamefont {J.}~\bibnamefont
  {Wu~Zhang}},\ }\bibfield  {title} {\bibinfo {title} {Mesoscale shallow
  convection in the atmosphere},\ }\href@noop {} {\bibfield  {journal}
  {\bibinfo  {journal} {Rev. Geophys.}\ }\textbf {\bibinfo {volume} {34}},\
  \bibinfo {pages} {403} (\bibinfo {year} {1996})}\BibitemShut {NoStop}%
\bibitem [{\citenamefont {Stevens}\ \emph {et~al.}(2018)\citenamefont
  {Stevens}, \citenamefont {Blass}, \citenamefont {Zhu}, \citenamefont
  {Verzicco},\ and\ \citenamefont {Lohse}}]{stevens2018turbulent}%
  \BibitemOpen
  \bibfield  {author} {\bibinfo {author} {\bibfnamefont {R.~J.}\ \bibnamefont
  {Stevens}}, \bibinfo {author} {\bibfnamefont {A.}~\bibnamefont {Blass}},
  \bibinfo {author} {\bibfnamefont {X.}~\bibnamefont {Zhu}}, \bibinfo {author}
  {\bibfnamefont {R.}~\bibnamefont {Verzicco}},\ and\ \bibinfo {author}
  {\bibfnamefont {D.}~\bibnamefont {Lohse}},\ }\bibfield  {title} {\bibinfo
  {title} {Turbulent thermal superstructures in {R}ayleigh-{B}{\'e}nard
  convection},\ }\href@noop {} {\bibfield  {journal} {\bibinfo  {journal}
  {Phys. Rev. Fluids}\ }\textbf {\bibinfo {volume} {3}},\ \bibinfo {pages}
  {041501(R)} (\bibinfo {year} {2018})}\BibitemShut {NoStop}%
\bibitem [{\citenamefont {Pandey}\ \emph {et~al.}(2018)\citenamefont {Pandey},
  \citenamefont {Scheel},\ and\ \citenamefont
  {Schumacher}}]{pandey2018turbulent}%
  \BibitemOpen
  \bibfield  {author} {\bibinfo {author} {\bibfnamefont {A.}~\bibnamefont
  {Pandey}}, \bibinfo {author} {\bibfnamefont {J.~D.}\ \bibnamefont {Scheel}},\
  and\ \bibinfo {author} {\bibfnamefont {J.}~\bibnamefont {Schumacher}},\
  }\bibfield  {title} {\bibinfo {title} {Turbulent superstructures in
  {R}ayleigh-{B}{\'e}nard convection},\ }\href@noop {} {\bibfield  {journal}
  {\bibinfo  {journal} {Nat. Commun.}\ }\textbf {\bibinfo {volume} {9}},\
  \bibinfo {pages} {1} (\bibinfo {year} {2018})}\BibitemShut {NoStop}%
\bibitem [{\citenamefont {Reddy}\ \emph {et~al.}(2016)\citenamefont {Reddy},
  \citenamefont {Celani}, \citenamefont {Sejnowski},\ and\ \citenamefont
  {Vergassola}}]{reddy2016learning}%
  \BibitemOpen
  \bibfield  {author} {\bibinfo {author} {\bibfnamefont {G.}~\bibnamefont
  {Reddy}}, \bibinfo {author} {\bibfnamefont {A.}~\bibnamefont {Celani}},
  \bibinfo {author} {\bibfnamefont {T.~J.}\ \bibnamefont {Sejnowski}},\ and\
  \bibinfo {author} {\bibfnamefont {M.}~\bibnamefont {Vergassola}},\ }\bibfield
   {title} {\bibinfo {title} {Learning to soar in turbulent environments},\
  }\href@noop {} {\bibfield  {journal} {\bibinfo  {journal} {Proc. Natl. Acad.
  Sci. U.S.A.}\ }\textbf {\bibinfo {volume} {113}},\ \bibinfo {pages} {4877}
  (\bibinfo {year} {2016})}\BibitemShut {NoStop}%
\bibitem [{\citenamefont {{\'A}kos}\ \emph {et~al.}(2008)\citenamefont
  {{\'A}kos}, \citenamefont {Nagy},\ and\ \citenamefont
  {Vicsek}}]{akos2008comparing}%
  \BibitemOpen
  \bibfield  {author} {\bibinfo {author} {\bibfnamefont {Z.}~\bibnamefont
  {{\'A}kos}}, \bibinfo {author} {\bibfnamefont {M.}~\bibnamefont {Nagy}},\
  and\ \bibinfo {author} {\bibfnamefont {T.}~\bibnamefont {Vicsek}},\
  }\bibfield  {title} {\bibinfo {title} {Comparing bird and human soaring
  strategies},\ }\href@noop {} {\bibfield  {journal} {\bibinfo  {journal}
  {Proc. Natl. Acad. Sci. U.S.A.}\ }\textbf {\bibinfo {volume} {105}},\
  \bibinfo {pages} {4139} (\bibinfo {year} {2008})}\BibitemShut {NoStop}%
\bibitem [{\citenamefont {Reddy}\ \emph {et~al.}(2018)\citenamefont {Reddy},
  \citenamefont {Wong-Ng}, \citenamefont {Celani}, \citenamefont {Sejnowski},\
  and\ \citenamefont {Vergassola}}]{reddy2018glider}%
  \BibitemOpen
  \bibfield  {author} {\bibinfo {author} {\bibfnamefont {G.}~\bibnamefont
  {Reddy}}, \bibinfo {author} {\bibfnamefont {J.}~\bibnamefont {Wong-Ng}},
  \bibinfo {author} {\bibfnamefont {A.}~\bibnamefont {Celani}}, \bibinfo
  {author} {\bibfnamefont {T.~J.}\ \bibnamefont {Sejnowski}},\ and\ \bibinfo
  {author} {\bibfnamefont {M.}~\bibnamefont {Vergassola}},\ }\bibfield  {title}
  {\bibinfo {title} {Glider soaring via reinforcement learning in the field},\
  }\href@noop {} {\bibfield  {journal} {\bibinfo  {journal} {Nature}\ }\textbf
  {\bibinfo {volume} {562}},\ \bibinfo {pages} {236} (\bibinfo {year}
  {2018})}\BibitemShut {NoStop}%
\bibitem [{\citenamefont {Xu}\ \emph {et~al.}(2022{\natexlab{a}})\citenamefont
  {Xu}, \citenamefont {Wu},\ and\ \citenamefont {Xi}}]{xu2022migration}%
  \BibitemOpen
  \bibfield  {author} {\bibinfo {author} {\bibfnamefont {A.}~\bibnamefont
  {Xu}}, \bibinfo {author} {\bibfnamefont {H.-L.}\ \bibnamefont {Wu}},\ and\
  \bibinfo {author} {\bibfnamefont {H.-D.}\ \bibnamefont {Xi}},\ }\bibfield
  {title} {\bibinfo {title} {Migration of self-propelling agent in a turbulent
  environment with minimal energy consumption},\ }\href@noop {} {\bibfield
  {journal} {\bibinfo  {journal} {Phys. Fluids}\ }\textbf {\bibinfo {volume}
  {34}},\ \bibinfo {pages} {035117} (\bibinfo {year}
  {2022}{\natexlab{a}})}\BibitemShut {NoStop}%
\bibitem [{\citenamefont {Xu}\ \emph {et~al.}(2017{\natexlab{a}})\citenamefont
  {Xu}, \citenamefont {Shyy},\ and\ \citenamefont {Zhao}}]{xu2017lattice}%
  \BibitemOpen
  \bibfield  {author} {\bibinfo {author} {\bibfnamefont {A.}~\bibnamefont
  {Xu}}, \bibinfo {author} {\bibfnamefont {W.}~\bibnamefont {Shyy}},\ and\
  \bibinfo {author} {\bibfnamefont {T.}~\bibnamefont {Zhao}},\ }\bibfield
  {title} {\bibinfo {title} {{L}attice-{B}oltzmann modeling of transport
  phenomena in fuel cells and flow batteries},\ }\href@noop {} {\bibfield
  {journal} {\bibinfo  {journal} {Acta Mech. Sin.}\ }\textbf {\bibinfo {volume}
  {33}},\ \bibinfo {pages} {555} (\bibinfo {year}
  {2017}{\natexlab{a}})}\BibitemShut {NoStop}%
\bibitem [{\citenamefont {Guo}\ and\ \citenamefont
  {Zheng}(2008)}]{guo2008analysis}%
  \BibitemOpen
  \bibfield  {author} {\bibinfo {author} {\bibfnamefont {Z.}~\bibnamefont
  {Guo}}\ and\ \bibinfo {author} {\bibfnamefont {C.}~\bibnamefont {Zheng}},\
  }\bibfield  {title} {\bibinfo {title} {Analysis of {L}attice-{B}oltzmann
  equation for microscale gas flows: {R}elaxation times, boundary conditions
  and the {K}nudsen layer},\ }\href@noop {} {\bibfield  {journal} {\bibinfo
  {journal} {Int. J. Comput. Fluid Dyn.}\ }\textbf {\bibinfo {volume} {22}},\
  \bibinfo {pages} {465} (\bibinfo {year} {2008})}\BibitemShut {NoStop}%
\bibitem [{\citenamefont {Xu}\ \emph {et~al.}(2017{\natexlab{b}})\citenamefont
  {Xu}, \citenamefont {Shi},\ and\ \citenamefont {Zhao}}]{xu2017accelerated}%
  \BibitemOpen
  \bibfield  {author} {\bibinfo {author} {\bibfnamefont {A.}~\bibnamefont
  {Xu}}, \bibinfo {author} {\bibfnamefont {L.}~\bibnamefont {Shi}},\ and\
  \bibinfo {author} {\bibfnamefont {T.}~\bibnamefont {Zhao}},\ }\bibfield
  {title} {\bibinfo {title} {Accelerated {L}attice-{B}oltzmann simulation using
  {GPU} and {O}pen{ACC} with data management},\ }\href@noop {} {\bibfield
  {journal} {\bibinfo  {journal} {Int. J. Heat Mass Transf.}\ }\textbf
  {\bibinfo {volume} {109}},\ \bibinfo {pages} {577} (\bibinfo {year}
  {2017}{\natexlab{b}})}\BibitemShut {NoStop}%
\bibitem [{\citenamefont {Xu}\ \emph {et~al.}(2019)\citenamefont {Xu},
  \citenamefont {Shi},\ and\ \citenamefont {Xi}}]{xu2019lattice}%
  \BibitemOpen
  \bibfield  {author} {\bibinfo {author} {\bibfnamefont {A.}~\bibnamefont
  {Xu}}, \bibinfo {author} {\bibfnamefont {L.}~\bibnamefont {Shi}},\ and\
  \bibinfo {author} {\bibfnamefont {H.-D.}\ \bibnamefont {Xi}},\ }\bibfield
  {title} {\bibinfo {title} {{L}attice-{B}oltzmann simulations of
  three-dimensional thermal convective flows at high {R}ayleigh number},\
  }\href@noop {} {\bibfield  {journal} {\bibinfo  {journal} {Int. J. Heat Mass
  Transf.}\ }\textbf {\bibinfo {volume} {140}},\ \bibinfo {pages} {359}
  (\bibinfo {year} {2019})}\BibitemShut {NoStop}%
\bibitem [{\citenamefont {Xu}\ and\ \citenamefont {Li}(2023)}]{xu2023multi}%
  \BibitemOpen
  \bibfield  {author} {\bibinfo {author} {\bibfnamefont {A.}~\bibnamefont
  {Xu}}\ and\ \bibinfo {author} {\bibfnamefont {B.-T.}\ \bibnamefont {Li}},\
  }\bibfield  {title} {\bibinfo {title} {{Multi-GPU thermal lattice Boltzmann
  simulations using OpenACC and MPI}},\ }\href@noop {} {\bibfield  {journal}
  {\bibinfo  {journal} {Int. J. Heat Mass Transf.}\ }\textbf {\bibinfo {volume}
  {201}},\ \bibinfo {pages} {123649} (\bibinfo {year} {2023})}\BibitemShut
  {NoStop}%
\bibitem [{\citenamefont {Castaing}\ \emph {et~al.}(1989)\citenamefont
  {Castaing}, \citenamefont {Gunaratne}, \citenamefont {Heslot}, \citenamefont
  {Kadanoff}, \citenamefont {Libchaber}, \citenamefont {Thomae}, \citenamefont
  {Wu}, \citenamefont {Zaleski},\ and\ \citenamefont
  {Zanetti}}]{castaing1989scaling}%
  \BibitemOpen
  \bibfield  {author} {\bibinfo {author} {\bibfnamefont {B.}~\bibnamefont
  {Castaing}}, \bibinfo {author} {\bibfnamefont {G.}~\bibnamefont {Gunaratne}},
  \bibinfo {author} {\bibfnamefont {F.}~\bibnamefont {Heslot}}, \bibinfo
  {author} {\bibfnamefont {L.}~\bibnamefont {Kadanoff}}, \bibinfo {author}
  {\bibfnamefont {A.}~\bibnamefont {Libchaber}}, \bibinfo {author}
  {\bibfnamefont {S.}~\bibnamefont {Thomae}}, \bibinfo {author} {\bibfnamefont
  {X.-Z.}\ \bibnamefont {Wu}}, \bibinfo {author} {\bibfnamefont
  {S.}~\bibnamefont {Zaleski}},\ and\ \bibinfo {author} {\bibfnamefont
  {G.}~\bibnamefont {Zanetti}},\ }\bibfield  {title} {\bibinfo {title} {Scaling
  of hard thermal turbulence in {R}ayleigh-{B}{\'e}nard convection},\
  }\href@noop {} {\bibfield  {journal} {\bibinfo  {journal} {J. Fluid Mech.}\
  }\textbf {\bibinfo {volume} {204}},\ \bibinfo {pages} {1} (\bibinfo {year}
  {1989})}\BibitemShut {NoStop}%
\bibitem [{\citenamefont {Krishna}\ \emph {et~al.}(2022)\citenamefont
  {Krishna}, \citenamefont {Song},\ and\ \citenamefont
  {Brunton}}]{krishna2022finite}%
  \BibitemOpen
  \bibfield  {author} {\bibinfo {author} {\bibfnamefont {K.}~\bibnamefont
  {Krishna}}, \bibinfo {author} {\bibfnamefont {Z.}~\bibnamefont {Song}},\ and\
  \bibinfo {author} {\bibfnamefont {S.~L.}\ \bibnamefont {Brunton}},\
  }\bibfield  {title} {\bibinfo {title} {Finite-horizon, energy-efficient
  trajectories in unsteady flows},\ }\href@noop {} {\bibfield  {journal}
  {\bibinfo  {journal} {Proc. R. Soc. A-Math. Phys. Eng. Sci.}\ }\textbf
  {\bibinfo {volume} {478}},\ \bibinfo {pages} {20210255} (\bibinfo {year}
  {2022})}\BibitemShut {NoStop}%
\bibitem [{\citenamefont {Colabrese}\ \emph {et~al.}(2017)\citenamefont
  {Colabrese}, \citenamefont {Gustavsson}, \citenamefont {Celani},\ and\
  \citenamefont {Biferale}}]{colabrese2017flow}%
  \BibitemOpen
  \bibfield  {author} {\bibinfo {author} {\bibfnamefont {S.}~\bibnamefont
  {Colabrese}}, \bibinfo {author} {\bibfnamefont {K.}~\bibnamefont
  {Gustavsson}}, \bibinfo {author} {\bibfnamefont {A.}~\bibnamefont {Celani}},\
  and\ \bibinfo {author} {\bibfnamefont {L.}~\bibnamefont {Biferale}},\
  }\bibfield  {title} {\bibinfo {title} {Flow navigation by smart microswimmers
  via reinforcement learning},\ }\href@noop {} {\bibfield  {journal} {\bibinfo
  {journal} {Phys. Rev. Lett.}\ }\textbf {\bibinfo {volume} {118}},\ \bibinfo
  {pages} {158004} (\bibinfo {year} {2017})}\BibitemShut {NoStop}%
\bibitem [{\citenamefont {Colabrese}\ \emph {et~al.}(2018)\citenamefont
  {Colabrese}, \citenamefont {Gustavsson}, \citenamefont {Celani},\ and\
  \citenamefont {Biferale}}]{colabrese2018smart}%
  \BibitemOpen
  \bibfield  {author} {\bibinfo {author} {\bibfnamefont {S.}~\bibnamefont
  {Colabrese}}, \bibinfo {author} {\bibfnamefont {K.}~\bibnamefont
  {Gustavsson}}, \bibinfo {author} {\bibfnamefont {A.}~\bibnamefont {Celani}},\
  and\ \bibinfo {author} {\bibfnamefont {L.}~\bibnamefont {Biferale}},\
  }\bibfield  {title} {\bibinfo {title} {Smart inertial particles},\
  }\href@noop {} {\bibfield  {journal} {\bibinfo  {journal} {Phys. Rev.
  Fluids}\ }\textbf {\bibinfo {volume} {3}},\ \bibinfo {pages} {084301}
  (\bibinfo {year} {2018})}\BibitemShut {NoStop}%
\bibitem [{\citenamefont {Schneider}\ and\ \citenamefont
  {Stark}(2019)}]{schneider2019optimal}%
  \BibitemOpen
  \bibfield  {author} {\bibinfo {author} {\bibfnamefont {E.}~\bibnamefont
  {Schneider}}\ and\ \bibinfo {author} {\bibfnamefont {H.}~\bibnamefont
  {Stark}},\ }\bibfield  {title} {\bibinfo {title} {Optimal steering of a smart
  active particle},\ }\href@noop {} {\bibfield  {journal} {\bibinfo  {journal}
  {Europhys. Lett.}\ }\textbf {\bibinfo {volume} {127}},\ \bibinfo {pages}
  {64003} (\bibinfo {year} {2019})}\BibitemShut {NoStop}%
\bibitem [{\citenamefont {Alageshan}\ \emph {et~al.}(2020)\citenamefont
  {Alageshan}, \citenamefont {Verma}, \citenamefont {Bec},\ and\ \citenamefont
  {Pandit}}]{alageshan2020machine}%
  \BibitemOpen
  \bibfield  {author} {\bibinfo {author} {\bibfnamefont {J.~K.}\ \bibnamefont
  {Alageshan}}, \bibinfo {author} {\bibfnamefont {A.~K.}\ \bibnamefont
  {Verma}}, \bibinfo {author} {\bibfnamefont {J.}~\bibnamefont {Bec}},\ and\
  \bibinfo {author} {\bibfnamefont {R.}~\bibnamefont {Pandit}},\ }\bibfield
  {title} {\bibinfo {title} {Machine learning strategies for path-planning
  microswimmers in turbulent flows},\ }\href@noop {} {\bibfield  {journal}
  {\bibinfo  {journal} {Phys. Rev. E}\ }\textbf {\bibinfo {volume} {101}},\
  \bibinfo {pages} {043110} (\bibinfo {year} {2020})}\BibitemShut {NoStop}%
\bibitem [{\citenamefont {Borra}\ \emph {et~al.}(2022)\citenamefont {Borra},
  \citenamefont {Biferale}, \citenamefont {Cencini},\ and\ \citenamefont
  {Celani}}]{borra2022reinforcement}%
  \BibitemOpen
  \bibfield  {author} {\bibinfo {author} {\bibfnamefont {F.}~\bibnamefont
  {Borra}}, \bibinfo {author} {\bibfnamefont {L.}~\bibnamefont {Biferale}},
  \bibinfo {author} {\bibfnamefont {M.}~\bibnamefont {Cencini}},\ and\ \bibinfo
  {author} {\bibfnamefont {A.}~\bibnamefont {Celani}},\ }\bibfield  {title}
  {\bibinfo {title} {Reinforcement learning for pursuit and evasion of
  microswimmers at low {R}eynolds number},\ }\href@noop {} {\bibfield
  {journal} {\bibinfo  {journal} {Phys. Rev. Fluids}\ }\textbf {\bibinfo
  {volume} {7}},\ \bibinfo {pages} {023103} (\bibinfo {year}
  {2022})}\BibitemShut {NoStop}%
\bibitem [{\citenamefont {Novati}\ \emph {et~al.}(2019)\citenamefont {Novati},
  \citenamefont {Mahadevan},\ and\ \citenamefont
  {Koumoutsakos}}]{novati2019controlled}%
  \BibitemOpen
  \bibfield  {author} {\bibinfo {author} {\bibfnamefont {G.}~\bibnamefont
  {Novati}}, \bibinfo {author} {\bibfnamefont {L.}~\bibnamefont {Mahadevan}},\
  and\ \bibinfo {author} {\bibfnamefont {P.}~\bibnamefont {Koumoutsakos}},\
  }\bibfield  {title} {\bibinfo {title} {Controlled gliding and perching
  through deep-reinforcement-learning},\ }\href@noop {} {\bibfield  {journal}
  {\bibinfo  {journal} {Phys. Rev. Fluids}\ }\textbf {\bibinfo {volume} {4}},\
  \bibinfo {pages} {093902} (\bibinfo {year} {2019})}\BibitemShut {NoStop}%
\bibitem [{\citenamefont {Zou}\ \emph {et~al.}(2022)\citenamefont {Zou},
  \citenamefont {Liu}, \citenamefont {Young}, \citenamefont {Pak},\ and\
  \citenamefont {Tsang}}]{zou2022gait}%
  \BibitemOpen
  \bibfield  {author} {\bibinfo {author} {\bibfnamefont {Z.}~\bibnamefont
  {Zou}}, \bibinfo {author} {\bibfnamefont {Y.}~\bibnamefont {Liu}}, \bibinfo
  {author} {\bibfnamefont {Y.-N.}\ \bibnamefont {Young}}, \bibinfo {author}
  {\bibfnamefont {O.~S.}\ \bibnamefont {Pak}},\ and\ \bibinfo {author}
  {\bibfnamefont {A.~C.}\ \bibnamefont {Tsang}},\ }\bibfield  {title} {\bibinfo
  {title} {Gait switching and targeted navigation of microswimmers via deep
  reinforcement learning},\ }\href@noop {} {\bibfield  {journal} {\bibinfo
  {journal} {Commun. Phys.}\ }\textbf {\bibinfo {volume} {5}},\ \bibinfo
  {pages} {158} (\bibinfo {year} {2022})}\BibitemShut {NoStop}%
\bibitem [{\citenamefont {Biferale}\ \emph {et~al.}(2019)\citenamefont
  {Biferale}, \citenamefont {Bonaccorso}, \citenamefont {Buzzicotti},
  \citenamefont {Clark Di~Leoni},\ and\ \citenamefont
  {Gustavsson}}]{biferale2019zermelo}%
  \BibitemOpen
  \bibfield  {author} {\bibinfo {author} {\bibfnamefont {L.}~\bibnamefont
  {Biferale}}, \bibinfo {author} {\bibfnamefont {F.}~\bibnamefont
  {Bonaccorso}}, \bibinfo {author} {\bibfnamefont {M.}~\bibnamefont
  {Buzzicotti}}, \bibinfo {author} {\bibfnamefont {P.}~\bibnamefont {Clark
  Di~Leoni}},\ and\ \bibinfo {author} {\bibfnamefont {K.}~\bibnamefont
  {Gustavsson}},\ }\bibfield  {title} {\bibinfo {title} {Zermelo’s problem:
  {O}ptimal point-to-point navigation in 2{D} turbulent flows using
  reinforcement learning},\ }\href@noop {} {\bibfield  {journal} {\bibinfo
  {journal} {Chaos}\ }\textbf {\bibinfo {volume} {29}},\ \bibinfo {pages}
  {103138} (\bibinfo {year} {2019})}\BibitemShut {NoStop}%
\bibitem [{\citenamefont {Garnier}\ \emph {et~al.}(2021)\citenamefont
  {Garnier}, \citenamefont {Viquerat}, \citenamefont {Rabault}, \citenamefont
  {Larcher}, \citenamefont {Kuhnle},\ and\ \citenamefont
  {Hachem}}]{garnier2021review}%
  \BibitemOpen
  \bibfield  {author} {\bibinfo {author} {\bibfnamefont {P.}~\bibnamefont
  {Garnier}}, \bibinfo {author} {\bibfnamefont {J.}~\bibnamefont {Viquerat}},
  \bibinfo {author} {\bibfnamefont {J.}~\bibnamefont {Rabault}}, \bibinfo
  {author} {\bibfnamefont {A.}~\bibnamefont {Larcher}}, \bibinfo {author}
  {\bibfnamefont {A.}~\bibnamefont {Kuhnle}},\ and\ \bibinfo {author}
  {\bibfnamefont {E.}~\bibnamefont {Hachem}},\ }\bibfield  {title} {\bibinfo
  {title} {A review on deep reinforcement learning for fluid mechanics},\
  }\href@noop {} {\bibfield  {journal} {\bibinfo  {journal} {Comput. Fluids}\
  }\textbf {\bibinfo {volume} {225}},\ \bibinfo {pages} {104973} (\bibinfo
  {year} {2021})}\BibitemShut {NoStop}%
\bibitem [{\citenamefont {Sutton}\ and\ \citenamefont
  {Barto}(2018)}]{sutton2018reinforcement}%
  \BibitemOpen
  \bibfield  {author} {\bibinfo {author} {\bibfnamefont {R.~S.}\ \bibnamefont
  {Sutton}}\ and\ \bibinfo {author} {\bibfnamefont {A.~G.}\ \bibnamefont
  {Barto}},\ }\href@noop {} {\emph {\bibinfo {title} {{Reinforcement Learning:
  An Introduction}}}}\ (\bibinfo  {publisher} {MIT Press, Cambridge, MA},\
  \bibinfo {year} {2018})\BibitemShut {NoStop}%
\bibitem [{\citenamefont {Mehta}\ \emph {et~al.}(2019)\citenamefont {Mehta},
  \citenamefont {Bukov}, \citenamefont {Wang}, \citenamefont {Day},
  \citenamefont {Richardson}, \citenamefont {Fisher},\ and\ \citenamefont
  {Schwab}}]{mehta2019high}%
  \BibitemOpen
  \bibfield  {author} {\bibinfo {author} {\bibfnamefont {P.}~\bibnamefont
  {Mehta}}, \bibinfo {author} {\bibfnamefont {M.}~\bibnamefont {Bukov}},
  \bibinfo {author} {\bibfnamefont {C.-H.}\ \bibnamefont {Wang}}, \bibinfo
  {author} {\bibfnamefont {A.~G.}\ \bibnamefont {Day}}, \bibinfo {author}
  {\bibfnamefont {C.}~\bibnamefont {Richardson}}, \bibinfo {author}
  {\bibfnamefont {C.~K.}\ \bibnamefont {Fisher}},\ and\ \bibinfo {author}
  {\bibfnamefont {D.~J.}\ \bibnamefont {Schwab}},\ }\bibfield  {title}
  {\bibinfo {title} {A high-bias, low-variance introduction to machine learning
  for physicists},\ }\href@noop {} {\bibfield  {journal} {\bibinfo  {journal}
  {Phys. Rep.}\ }\textbf {\bibinfo {volume} {810}},\ \bibinfo {pages} {1}
  (\bibinfo {year} {2019})}\BibitemShut {NoStop}%
\bibitem [{\citenamefont {Brunton}\ \emph {et~al.}(2020)\citenamefont
  {Brunton}, \citenamefont {Noack},\ and\ \citenamefont
  {Koumoutsakos}}]{brunton2020machine}%
  \BibitemOpen
  \bibfield  {author} {\bibinfo {author} {\bibfnamefont {S.~L.}\ \bibnamefont
  {Brunton}}, \bibinfo {author} {\bibfnamefont {B.~R.}\ \bibnamefont {Noack}},\
  and\ \bibinfo {author} {\bibfnamefont {P.}~\bibnamefont {Koumoutsakos}},\
  }\bibfield  {title} {\bibinfo {title} {Machine learning for fluid
  mechanics},\ }\href@noop {} {\bibfield  {journal} {\bibinfo  {journal} {Annu.
  Rev. Fluid Mech.}\ }\textbf {\bibinfo {volume} {52}},\ \bibinfo {pages} {477}
  (\bibinfo {year} {2020})}\BibitemShut {NoStop}%
\bibitem [{\citenamefont {Cichos}\ \emph {et~al.}(2020)\citenamefont {Cichos},
  \citenamefont {Gustavsson}, \citenamefont {Mehlig},\ and\ \citenamefont
  {Volpe}}]{cichos2020machine}%
  \BibitemOpen
  \bibfield  {author} {\bibinfo {author} {\bibfnamefont {F.}~\bibnamefont
  {Cichos}}, \bibinfo {author} {\bibfnamefont {K.}~\bibnamefont {Gustavsson}},
  \bibinfo {author} {\bibfnamefont {B.}~\bibnamefont {Mehlig}},\ and\ \bibinfo
  {author} {\bibfnamefont {G.}~\bibnamefont {Volpe}},\ }\bibfield  {title}
  {\bibinfo {title} {Machine learning for active matter},\ }\href@noop {}
  {\bibfield  {journal} {\bibinfo  {journal} {Nat. Mach. Intell.}\ }\textbf
  {\bibinfo {volume} {2}},\ \bibinfo {pages} {94} (\bibinfo {year}
  {2020})}\BibitemShut {NoStop}%
\bibitem [{\citenamefont {Tsang}\ \emph {et~al.}(2020)\citenamefont {Tsang},
  \citenamefont {Tong}, \citenamefont {Nallan},\ and\ \citenamefont
  {Pak}}]{tsang2020self}%
  \BibitemOpen
  \bibfield  {author} {\bibinfo {author} {\bibfnamefont {A.~C.~H.}\
  \bibnamefont {Tsang}}, \bibinfo {author} {\bibfnamefont {P.~W.}\ \bibnamefont
  {Tong}}, \bibinfo {author} {\bibfnamefont {S.}~\bibnamefont {Nallan}},\ and\
  \bibinfo {author} {\bibfnamefont {O.~S.}\ \bibnamefont {Pak}},\ }\bibfield
  {title} {\bibinfo {title} {Self-learning how to swim at low {R}eynolds
  number},\ }\href@noop {} {\bibfield  {journal} {\bibinfo  {journal} {Phys.
  Rev. Fluids}\ }\textbf {\bibinfo {volume} {5}},\ \bibinfo {pages} {074101}
  (\bibinfo {year} {2020})}\BibitemShut {NoStop}%
\bibitem [{\citenamefont {Mui{\~n}os-Landin}\ \emph {et~al.}(2021)\citenamefont
  {Mui{\~n}os-Landin}, \citenamefont {Fischer}, \citenamefont {Holubec},\ and\
  \citenamefont {Cichos}}]{muinos2021reinforcement}%
  \BibitemOpen
  \bibfield  {author} {\bibinfo {author} {\bibfnamefont {S.}~\bibnamefont
  {Mui{\~n}os-Landin}}, \bibinfo {author} {\bibfnamefont {A.}~\bibnamefont
  {Fischer}}, \bibinfo {author} {\bibfnamefont {V.}~\bibnamefont {Holubec}},\
  and\ \bibinfo {author} {\bibfnamefont {F.}~\bibnamefont {Cichos}},\
  }\bibfield  {title} {\bibinfo {title} {Reinforcement learning with artificial
  microswimmers},\ }\href@noop {} {\bibfield  {journal} {\bibinfo  {journal}
  {Sci. Robot.}\ }\textbf {\bibinfo {volume} {6}},\ \bibinfo {pages} {eabd9285}
  (\bibinfo {year} {2021})}\BibitemShut {NoStop}%
\bibitem [{\citenamefont {Monderkamp}\ \emph {et~al.}(2022)\citenamefont
  {Monderkamp}, \citenamefont {Schwarzendahl}, \citenamefont {Klatt},\ and\
  \citenamefont {L{\"o}wen}}]{monderkamp2022active}%
  \BibitemOpen
  \bibfield  {author} {\bibinfo {author} {\bibfnamefont {P.~A.}\ \bibnamefont
  {Monderkamp}}, \bibinfo {author} {\bibfnamefont {F.~J.}\ \bibnamefont
  {Schwarzendahl}}, \bibinfo {author} {\bibfnamefont {M.~A.}\ \bibnamefont
  {Klatt}},\ and\ \bibinfo {author} {\bibfnamefont {H.}~\bibnamefont
  {L{\"o}wen}},\ }\bibfield  {title} {\bibinfo {title} {Active particles using
  reinforcement learning to navigate in complex motility landscapes},\
  }\href@noop {} {\bibfield  {journal} {\bibinfo  {journal} {Mach. Learn.-Sci.
  Technol.}\ }\textbf {\bibinfo {volume} {3}},\ \bibinfo {pages} {045024}
  (\bibinfo {year} {2022})}\BibitemShut {NoStop}%
\bibitem [{\citenamefont {Gazzola}\ \emph {et~al.}(2016)\citenamefont
  {Gazzola}, \citenamefont {Tchieu}, \citenamefont {Alexeev}, \citenamefont
  {de~Brauer},\ and\ \citenamefont {Koumoutsakos}}]{gazzola2016learning}%
  \BibitemOpen
  \bibfield  {author} {\bibinfo {author} {\bibfnamefont {M.}~\bibnamefont
  {Gazzola}}, \bibinfo {author} {\bibfnamefont {A.~A.}\ \bibnamefont {Tchieu}},
  \bibinfo {author} {\bibfnamefont {D.}~\bibnamefont {Alexeev}}, \bibinfo
  {author} {\bibfnamefont {A.}~\bibnamefont {de~Brauer}},\ and\ \bibinfo
  {author} {\bibfnamefont {P.}~\bibnamefont {Koumoutsakos}},\ }\bibfield
  {title} {\bibinfo {title} {Learning to school in the presence of hydrodynamic
  interactions},\ }\href@noop {} {\bibfield  {journal} {\bibinfo  {journal} {J.
  Fluid Mech.}\ }\textbf {\bibinfo {volume} {789}},\ \bibinfo {pages} {726}
  (\bibinfo {year} {2016})}\BibitemShut {NoStop}%
\bibitem [{\citenamefont {Verma}\ \emph {et~al.}(2018)\citenamefont {Verma},
  \citenamefont {Novati},\ and\ \citenamefont
  {Koumoutsakos}}]{verma2018efficient}%
  \BibitemOpen
  \bibfield  {author} {\bibinfo {author} {\bibfnamefont {S.}~\bibnamefont
  {Verma}}, \bibinfo {author} {\bibfnamefont {G.}~\bibnamefont {Novati}},\ and\
  \bibinfo {author} {\bibfnamefont {P.}~\bibnamefont {Koumoutsakos}},\
  }\bibfield  {title} {\bibinfo {title} {Efficient collective swimming by
  harnessing vortices through deep reinforcement learning},\ }\href@noop {}
  {\bibfield  {journal} {\bibinfo  {journal} {Proc. Natl. Acad. Sci. U.S.A.}\
  }\textbf {\bibinfo {volume} {115}},\ \bibinfo {pages} {5849} (\bibinfo {year}
  {2018})}\BibitemShut {NoStop}%
\bibitem [{\citenamefont {Haarnoja}\ \emph {et~al.}(2018)\citenamefont
  {Haarnoja}, \citenamefont {Zhou}, \citenamefont {Abbeel},\ and\ \citenamefont
  {Levine}}]{haarnoja2018soft}%
  \BibitemOpen
  \bibfield  {author} {\bibinfo {author} {\bibfnamefont {T.}~\bibnamefont
  {Haarnoja}}, \bibinfo {author} {\bibfnamefont {A.}~\bibnamefont {Zhou}},
  \bibinfo {author} {\bibfnamefont {P.}~\bibnamefont {Abbeel}},\ and\ \bibinfo
  {author} {\bibfnamefont {S.}~\bibnamefont {Levine}},\ }\bibfield  {title}
  {\bibinfo {title} {Soft actor-critic: {O}ff-policy maximum entropy deep
  reinforcement learning with a stochastic actor},\ }in\ \href@noop {} {\emph
  {\bibinfo {booktitle} {Proceedings of the International Conference on Machine
  Learning}}}\ (\bibinfo {organization} {PMLR},\ \bibinfo {year} {2018})\ pp.\
  \bibinfo {pages} {1861--1870}\BibitemShut {NoStop}%
\bibitem [{\citenamefont {Wang}\ \emph
  {et~al.}(2020{\natexlab{a}})\citenamefont {Wang}, \citenamefont {Verzicco},
  \citenamefont {Lohse},\ and\ \citenamefont {Shishkina}}]{wang2020multiple}%
  \BibitemOpen
  \bibfield  {author} {\bibinfo {author} {\bibfnamefont {Q.}~\bibnamefont
  {Wang}}, \bibinfo {author} {\bibfnamefont {R.}~\bibnamefont {Verzicco}},
  \bibinfo {author} {\bibfnamefont {D.}~\bibnamefont {Lohse}},\ and\ \bibinfo
  {author} {\bibfnamefont {O.}~\bibnamefont {Shishkina}},\ }\bibfield  {title}
  {\bibinfo {title} {Multiple states in turbulent large-aspect-ratio thermal
  convection: {W}hat determines the number of convection rolls?},\ }\href@noop
  {} {\bibfield  {journal} {\bibinfo  {journal} {Phys. Rev. Lett.}\ }\textbf
  {\bibinfo {volume} {125}},\ \bibinfo {pages} {074501} (\bibinfo {year}
  {2020}{\natexlab{a}})}\BibitemShut {NoStop}%
\bibitem [{\citenamefont {Xu}\ \emph {et~al.}(2020)\citenamefont {Xu},
  \citenamefont {Chen}, \citenamefont {Wang},\ and\ \citenamefont
  {Xi}}]{xu2020correlation}%
  \BibitemOpen
  \bibfield  {author} {\bibinfo {author} {\bibfnamefont {A.}~\bibnamefont
  {Xu}}, \bibinfo {author} {\bibfnamefont {X.}~\bibnamefont {Chen}}, \bibinfo
  {author} {\bibfnamefont {F.}~\bibnamefont {Wang}},\ and\ \bibinfo {author}
  {\bibfnamefont {H.-D.}\ \bibnamefont {Xi}},\ }\bibfield  {title} {\bibinfo
  {title} {Correlation of internal flow structure with heat transfer efficiency
  in turbulent {R}ayleigh-{B}{\'e}nard convection},\ }\href@noop {} {\bibfield
  {journal} {\bibinfo  {journal} {Phys. Fluids}\ }\textbf {\bibinfo {volume}
  {32}},\ \bibinfo {pages} {105112} (\bibinfo {year} {2020})}\BibitemShut
  {NoStop}%
\bibitem [{\citenamefont {Halko}\ \emph {et~al.}(2011)\citenamefont {Halko},
  \citenamefont {Martinsson},\ and\ \citenamefont {Tropp}}]{halko2011finding}%
  \BibitemOpen
  \bibfield  {author} {\bibinfo {author} {\bibfnamefont {N.}~\bibnamefont
  {Halko}}, \bibinfo {author} {\bibfnamefont {P.-G.}\ \bibnamefont
  {Martinsson}},\ and\ \bibinfo {author} {\bibfnamefont {J.~A.}\ \bibnamefont
  {Tropp}},\ }\bibfield  {title} {\bibinfo {title} {Finding structure with
  randomness: {P}robabilistic algorithms for constructing approximate matrix
  decompositions},\ }\href@noop {} {\bibfield  {journal} {\bibinfo  {journal}
  {SIAM Rev.}\ }\textbf {\bibinfo {volume} {53}},\ \bibinfo {pages} {217}
  (\bibinfo {year} {2011})}\BibitemShut {NoStop}%
\bibitem [{\citenamefont {Zhang}\ \emph {et~al.}(2017)\citenamefont {Zhang},
  \citenamefont {Zhou},\ and\ \citenamefont {Sun}}]{zhang2017statistics}%
  \BibitemOpen
  \bibfield  {author} {\bibinfo {author} {\bibfnamefont {Y.}~\bibnamefont
  {Zhang}}, \bibinfo {author} {\bibfnamefont {Q.}~\bibnamefont {Zhou}},\ and\
  \bibinfo {author} {\bibfnamefont {C.}~\bibnamefont {Sun}},\ }\bibfield
  {title} {\bibinfo {title} {{Statistics of kinetic and thermal energy
  dissipation rates in two-dimensional turbulent Rayleigh--B{\'e}nard
  convection}},\ }\href@noop {} {\bibfield  {journal} {\bibinfo  {journal} {J.
  Fluid Mech.}\ }\textbf {\bibinfo {volume} {814}},\ \bibinfo {pages} {165}
  (\bibinfo {year} {2017})}\BibitemShut {NoStop}%
\bibitem [{\citenamefont {Zhu}\ \emph {et~al.}(2018)\citenamefont {Zhu},
  \citenamefont {Mathai}, \citenamefont {Stevens}, \citenamefont {Verzicco},\
  and\ \citenamefont {Lohse}}]{zhu2018transition}%
  \BibitemOpen
  \bibfield  {author} {\bibinfo {author} {\bibfnamefont {X.}~\bibnamefont
  {Zhu}}, \bibinfo {author} {\bibfnamefont {V.}~\bibnamefont {Mathai}},
  \bibinfo {author} {\bibfnamefont {R.~J.}\ \bibnamefont {Stevens}}, \bibinfo
  {author} {\bibfnamefont {R.}~\bibnamefont {Verzicco}},\ and\ \bibinfo
  {author} {\bibfnamefont {D.}~\bibnamefont {Lohse}},\ }\bibfield  {title}
  {\bibinfo {title} {{Transition to the ultimate regime in two-dimensional
  Rayleigh-B{\'e}nard convection}},\ }\href@noop {} {\bibfield  {journal}
  {\bibinfo  {journal} {Phys. Rev. Lett.}\ }\textbf {\bibinfo {volume} {120}},\
  \bibinfo {pages} {144502} (\bibinfo {year} {2018})}\BibitemShut {NoStop}%
\bibitem [{\citenamefont {Wang}\ \emph
  {et~al.}(2020{\natexlab{b}})\citenamefont {Wang}, \citenamefont {Zhou},\ and\
  \citenamefont {Sun}}]{wang2020vibration}%
  \BibitemOpen
  \bibfield  {author} {\bibinfo {author} {\bibfnamefont {B.-F.}\ \bibnamefont
  {Wang}}, \bibinfo {author} {\bibfnamefont {Q.}~\bibnamefont {Zhou}},\ and\
  \bibinfo {author} {\bibfnamefont {C.}~\bibnamefont {Sun}},\ }\bibfield
  {title} {\bibinfo {title} {Vibration-induced boundary-layer destabilization
  achieves massive heat-transport enhancement},\ }\href@noop {} {\bibfield
  {journal} {\bibinfo  {journal} {Sci. Adv.}\ }\textbf {\bibinfo {volume}
  {6}},\ \bibinfo {pages} {eaaz8239} (\bibinfo {year}
  {2020}{\natexlab{b}})}\BibitemShut {NoStop}%
\bibitem [{\citenamefont {Chen}\ \emph {et~al.}(2019)\citenamefont {Chen},
  \citenamefont {Huang}, \citenamefont {Xia},\ and\ \citenamefont
  {Xi}}]{chen2019emergence}%
  \BibitemOpen
  \bibfield  {author} {\bibinfo {author} {\bibfnamefont {X.}~\bibnamefont
  {Chen}}, \bibinfo {author} {\bibfnamefont {S.-D.}\ \bibnamefont {Huang}},
  \bibinfo {author} {\bibfnamefont {K.-Q.}\ \bibnamefont {Xia}},\ and\ \bibinfo
  {author} {\bibfnamefont {H.-D.}\ \bibnamefont {Xi}},\ }\bibfield  {title}
  {\bibinfo {title} {Emergence of substructures inside the large-scale
  circulation induces transition in flow reversals in turbulent thermal
  convection},\ }\href@noop {} {\bibfield  {journal} {\bibinfo  {journal} {J.
  Fluid Mech.}\ }\textbf {\bibinfo {volume} {877}},\ \bibinfo {pages} {R1}
  (\bibinfo {year} {2019})}\BibitemShut {NoStop}%
\bibitem [{SM()}]{SM}%
  \BibitemOpen
  \bibfield  {title} {\bibinfo {title} {See supplemental material at
  http://link.aps.org/supplemental/10.1103/physrevfluids.8.023502 for
  comparison of the smart agent and the naive agent migrating in turbulent
  {R}ayleigh-{B}{\'e}nard convection},\ }\href@noop {} {\ }\BibitemShut
  {NoStop}%
\bibitem [{\citenamefont {Gunnarson}\ \emph {et~al.}(2021)\citenamefont
  {Gunnarson}, \citenamefont {Mandralis}, \citenamefont {Novati}, \citenamefont
  {Koumoutsakos},\ and\ \citenamefont {Dabiri}}]{gunnarson2021learning}%
  \BibitemOpen
  \bibfield  {author} {\bibinfo {author} {\bibfnamefont {P.}~\bibnamefont
  {Gunnarson}}, \bibinfo {author} {\bibfnamefont {I.}~\bibnamefont
  {Mandralis}}, \bibinfo {author} {\bibfnamefont {G.}~\bibnamefont {Novati}},
  \bibinfo {author} {\bibfnamefont {P.}~\bibnamefont {Koumoutsakos}},\ and\
  \bibinfo {author} {\bibfnamefont {J.~O.}\ \bibnamefont {Dabiri}},\ }\bibfield
   {title} {\bibinfo {title} {Learning efficient navigation in vortical flow
  fields},\ }\href@noop {} {\bibfield  {journal} {\bibinfo  {journal} {Nat.
  Commun.}\ }\textbf {\bibinfo {volume} {12}},\ \bibinfo {pages} {7143}
  (\bibinfo {year} {2021})}\BibitemShut {NoStop}%
\bibitem [{\citenamefont {Gustavsson}\ \emph {et~al.}(2017)\citenamefont
  {Gustavsson}, \citenamefont {Biferale}, \citenamefont {Celani},\ and\
  \citenamefont {Colabrese}}]{gustavsson2017finding}%
  \BibitemOpen
  \bibfield  {author} {\bibinfo {author} {\bibfnamefont {K.}~\bibnamefont
  {Gustavsson}}, \bibinfo {author} {\bibfnamefont {L.}~\bibnamefont
  {Biferale}}, \bibinfo {author} {\bibfnamefont {A.}~\bibnamefont {Celani}},\
  and\ \bibinfo {author} {\bibfnamefont {S.}~\bibnamefont {Colabrese}},\
  }\bibfield  {title} {\bibinfo {title} {Finding efficient swimming strategies
  in a three-dimensional chaotic flow by reinforcement learning},\ }\href@noop
  {} {\bibfield  {journal} {\bibinfo  {journal} {Eur. Phys. J. E}\ }\textbf
  {\bibinfo {volume} {40}},\ \bibinfo {pages} {110} (\bibinfo {year}
  {2017})}\BibitemShut {NoStop}%
\bibitem [{\citenamefont {Kubo}\ and\ \citenamefont
  {Shimizu}(2022)}]{kubo2022efficient}%
  \BibitemOpen
  \bibfield  {author} {\bibinfo {author} {\bibfnamefont {A.}~\bibnamefont
  {Kubo}}\ and\ \bibinfo {author} {\bibfnamefont {M.}~\bibnamefont {Shimizu}},\
  }\bibfield  {title} {\bibinfo {title} {Efficient reinforcement learning with
  partial observables for fluid flow control},\ }\href@noop {} {\bibfield
  {journal} {\bibinfo  {journal} {Phys. Rev. E}\ }\textbf {\bibinfo {volume}
  {105}},\ \bibinfo {pages} {065101} (\bibinfo {year} {2022})}\BibitemShut
  {NoStop}%
\bibitem [{\citenamefont {Qiu}\ \emph {et~al.}(2022{\natexlab{a}})\citenamefont
  {Qiu}, \citenamefont {Mousavi}, \citenamefont {Gustavsson}, \citenamefont
  {Xu}, \citenamefont {Mehlig},\ and\ \citenamefont
  {Zhao}}]{qiu2022navigation}%
  \BibitemOpen
  \bibfield  {author} {\bibinfo {author} {\bibfnamefont {J.}~\bibnamefont
  {Qiu}}, \bibinfo {author} {\bibfnamefont {N.}~\bibnamefont {Mousavi}},
  \bibinfo {author} {\bibfnamefont {K.}~\bibnamefont {Gustavsson}}, \bibinfo
  {author} {\bibfnamefont {C.}~\bibnamefont {Xu}}, \bibinfo {author}
  {\bibfnamefont {B.}~\bibnamefont {Mehlig}},\ and\ \bibinfo {author}
  {\bibfnamefont {L.}~\bibnamefont {Zhao}},\ }\bibfield  {title} {\bibinfo
  {title} {Navigation of micro-swimmers in steady flow: {T}he importance of
  symmetries},\ }\href@noop {} {\bibfield  {journal} {\bibinfo  {journal} {J.
  Fluid Mech.}\ }\textbf {\bibinfo {volume} {932}},\ \bibinfo {pages} {A10}
  (\bibinfo {year} {2022}{\natexlab{a}})}\BibitemShut {NoStop}%
\bibitem [{\citenamefont {Qiu}\ \emph {et~al.}(2022{\natexlab{b}})\citenamefont
  {Qiu}, \citenamefont {Mousavi}, \citenamefont {Zhao},\ and\ \citenamefont
  {Gustavsson}}]{qiu2022active}%
  \BibitemOpen
  \bibfield  {author} {\bibinfo {author} {\bibfnamefont {J.}~\bibnamefont
  {Qiu}}, \bibinfo {author} {\bibfnamefont {N.}~\bibnamefont {Mousavi}},
  \bibinfo {author} {\bibfnamefont {L.}~\bibnamefont {Zhao}},\ and\ \bibinfo
  {author} {\bibfnamefont {K.}~\bibnamefont {Gustavsson}},\ }\bibfield  {title}
  {\bibinfo {title} {Active gyrotactic stability of microswimmers using
  hydromechanical signals},\ }\href@noop {} {\bibfield  {journal} {\bibinfo
  {journal} {Phys. Rev. Fluids}\ }\textbf {\bibinfo {volume} {7}},\ \bibinfo
  {pages} {014311} (\bibinfo {year} {2022}{\natexlab{b}})}\BibitemShut
  {NoStop}%
\bibitem [{\citenamefont {Xu}\ \emph {et~al.}(2022{\natexlab{b}})\citenamefont
  {Xu}, \citenamefont {Xu}, \citenamefont {Jiang},\ and\ \citenamefont
  {Xi}}]{xu2022production}%
  \BibitemOpen
  \bibfield  {author} {\bibinfo {author} {\bibfnamefont {A.}~\bibnamefont
  {Xu}}, \bibinfo {author} {\bibfnamefont {B.-R.}\ \bibnamefont {Xu}}, \bibinfo
  {author} {\bibfnamefont {L.-S.}\ \bibnamefont {Jiang}},\ and\ \bibinfo
  {author} {\bibfnamefont {H.-D.}\ \bibnamefont {Xi}},\ }\bibfield  {title}
  {\bibinfo {title} {Production and transport of vorticity in two-dimensional
  {R}ayleigh-{B}{\'e}nard convection cell},\ }\href@noop {} {\bibfield
  {journal} {\bibinfo  {journal} {Phys. Fluids}\ }\textbf {\bibinfo {volume}
  {34}},\ \bibinfo {pages} {013609} (\bibinfo {year}
  {2022}{\natexlab{b}})}\BibitemShut {NoStop}%
\end{thebibliography}
%

\end{document}